# Further analysis of metagenomic datasets containing GD and GX pangolin CoVs indicates widespread contamination, undermining pangolin host attribution

Adrian Jones[1], Steven E. Massey[2], Daoyu Zhang[3] , Yuri Deigin[4*] and Steven C. Quay[5]

[1] Independent Bioinformatics Researcher, Melbourne, Australia

[2] Biology Dept, University of Puerto Rico - Rio Piedras, PR USA

[3] Independent Genetics Researcher, Sydney, Australia

[4] Youthereum Genetics Inc., Toronto, Ontario, Canada; ORCID 0000-0002-3397-5811

[5] Atossa Therapeutics, Inc., Seattle, WA USA; ORCID 0000-0002-0363-7651

[*]Correspondence to: ydeigin@protonmail.com

## Abstract

The only animals other than bats reported to have been infected with SARS-CoV-2-related coronaviruses (SARS2r-CoVs) prior to the COVID-19 pandemic are pangolins. In early 2020 multiple papers reported the identification of two clades of SARS2r-CoVs, GD and GX, infecting pangolins. However the RNA-Seq datasets supporting pangolin genome assembly were widely contaminated, contained synthetic vectors or were heavily enriched or filtered with little but coronavirus sequences left in the datasets. Here we investigate two pangolin fecal samples sequenced by Li et al. (2021) provided in support of GD PCoV infection of pangolins in Guangdong and find the read distribution consistent with PCR amplicon contamination and SARS-CoV-2 contamination, and further identify the presence of synthetic plasmid sequences. We also build upon our previous work to further analyze the dataset GX/P3B by Lam et al. (2020), which is the only non enriched/heavily filtered pangolin tissue dataset sequenced by Lam et al. (2020). We identify synthetic vectors and confirm human genomic origin samples in the dataset. Finally, we find human mitochondrial sequences in all pangolin organ datasets and mouse and tiger mitochondrial sequences in selected pangolin organ datasets sequenced by Liu et al. (2019). We infer that human and mouse genomic origin sequences were probably sourced from contamination prior to sequencing, while tiger origin sequence contamination may have occurred due to index hopping during sequencing. These observations are problematic for attributing pangolins as SARS2r-CoV hosts in the datasets examined. The forensic methods developed and used here can be applied to examine any third party SRA data sets.

## Introduction

Identifying the origin of SARS-CoV-2 is a critical question which has implications for preventing a similar pandemic in the future. Although several bat coronavirus genomes have a higher identity to SARS-CoV-2 on a full genome level, pangolin CoVs MP789 and the nearly identical GD_1 genome have a 96.9% amino acid identity in the receptor binding domain (RBD) to SARS-CoV-2, but only a 90.57% full genome nucleotide identity (EPI_ISL_41072). The RBD in BANAL-20-52/Laos/2020 has one amino acid higher identity to SARS-CoV-2 than MP789/GD_1 (217/223aa), yielding for 97.31% amino acid identity and 96.85% full genome nucleotide identity.

Prior to the discovery of the Laotian BANAL CoVs, the high amino acid identity between the RBDs of MP789 and SARS-CoV-2 has led to multiple papers proposing recombination as the source of the RBD in SARS-CoV-2 (Wong et al. 2020; Andersen et al. 2020; Flores-Alanis et al. 2020; Piplani et al. 2021). An unanswered question is how pangolins came to be infected with a SARS2r-CoV with significant similarity to SARS-CoV-2. Furthermore, what is the evolutionary relationship between the RBDs of the BANAL CoVs, GD PCoVs and SARS-CoV-2, especially as the viruses are found in different host species.

Jones et al. (2022a) proposed that the SARS2r-CoV sequences found in the datasets supporting the assembly of partial MP789/GD_1 genomes sequenced by Liu et al. (2019) were not related to a primary infection of pangolins with a SARS2r-CoV, but likely due to inadvertent contamination of the datasets by a human cell or transgenic mouse model experiments. An additional BioProject by Xiao et al. (2020) in support of full assembly of the GD_1 genome was found to rely on GD_1 genome sequences inserted into synthetic plasmids, and circular read coverage consistent with molecular cloning was found. In addition, datasets supplied by Liu et al. (2020) to support the assembly of MP789 contained a 3nt gap in coverage and SNV that precludes a full assembly of the genome (Jones et al. 2022a).

Although raw sequencing datasets supporting the assembly of GD PCoVs are problematic, published datasets supporting Guangxi (GX) pangolin coronaviruses (PCoVs) are even more so. GX PCoVs were first identified by Lam et al. (2020), with a publication of 6 highly similar GX PCoVs identified in sequencing datasets from frozen *Manis javanica* tissue samples collected between August 2017 and January 2018 by the Guangxi Customs Bureau. All samples except one, GX/P3B, were either heavily filtered or enriched, with the absence of non-CoV reads making host verification and non-CoV contamination review impossible. Even with extensive filtering, sample GX/P2V was found to contain 20 reads with exact matches to the SARS-CoV-2 Wuhan-Hu-1 genome. The only non heavily filtered tissue sample, GX/P3B, is contaminated with human genomic origin material with 4.3% of mitochondrial aligned reads matching the *Homo sapiens* mitochondrial genome.

Peng et al. (2021) documented the discovery of a partial sequence of a GX PCoV-related virus MP20 (EPI_ISL_610156), sequenced from a *Manis pentadactyla* (MP) sampled in Yunnan province in 2017. Genomic DNA from the same Chinese pangolin MP20 was previously sequenced by Hu et al. (2020). Peng et al. (2021) propose PCoV MP20 supports the hypothesis that pangolins are natural hosts for CoVs and is the first documentation of MP as a host species for CoVs. However the release of raw sequencing data in support of PCoV MP20 assembly in BioProject PRJCA003816 on the National Genomics Data Center (NGDC) Genome Sequence Archive (GSA) database (https://ngdc.cncb.ac.cn/gsa/) has been embargoed until September 2023. Consequently, this claim cannot be validated.

Infection of *Manis javanica* pangolins in Hung Yen, Vietnam by a GX PCoV-related virus was documented by Nga et al. (2022) but only partial RdRp and NSP14 gene fragments were recovered and no RNA-Seq was conducted. In another GX PCoV discovery, Jones et al. (2022b) identified a novel GX PCoV most similar to GX/P2V, contaminating human cell Alphavirus infectivity experiments conducted at the WIV in 2021. A further GX-related CoV discovery was made by Jones et al. (2022c) where an intriguing GX-related CoV, GX_ZC45r-CoV was identified in game animal dataset sampled across China and sequenced by He et al. (2022). In a further analysis by Jones et al. (2022c), both the NSP4 and NSP10 coding regions of the novel CoV were found to group with the GX PCoV clade, but the RdRp phylogenetically sits with bat-SL-CoVZC45. The CoV is interpreted to stem from laboratory contamination with clustered read coverage and a partial synthetic vector sequence attached to a read indicating the CoV was sequenced from cDNA contained in plasmids. The host may be *Rhinolophus pusillus* bats located in Zhoushan city, Zhejiang Province which are the host of bat-SL-CoVZC45 (Hu et al. 2018).

Here we assess two RNA-Seq datasets of pangolin fecal samples DG14 and DG18 provided as part of an epidemiological study of Pangolin CoVs by Li et al. (2021). We find that DG14 and DG18 contain trace levels of a GD PCoV-related virus but only a very small fraction of the genome was recovered, in addition to contamination from human and mouse origin sequences. We further undertake additional analysis to Jones et al. (2022a) on the only non-enriched or heavily filtered pangolin specimen dataset in support of GX PCoVs, GX/P3B, showing plasmid and human genomic origin sequence contamination and several bacterial species inconsistent with a pangolin blood sample. Lastly we undertake additional mitochondrial genome screening and bacterial taxonomic analysis of datasets in Liu et al. (2019) and infer that human and mouse genomic origin content, as well as bacterial contamination, are likely related to upstream contamination, while tiger origin genomic material may be sourced from index-hopping during the sequencing run. At this time, no pangolin-derived BioProject containing SARS2r-CoV sequences can provide strong evidence that pangolins are a host for SARS2r-CoVs, as all such BioProjects either have significant genomic contamination and or have SRAs which have been heavily filtered or enriched, or have insufficient read coverage to allow full genome assembly (Jones et al. 2022a).

# Results

### DG14 and DG18: PCoV and SARS-CoV-2

Li et al. (2021) reported that three beta-CoV-positive *M. javanica* pangolins had been identified out of 43 captive pangolins sampled from Dongguan, Shenzhen, and Shaoguan in the Guangdong Province of China. Two fecal sample SRA datasets corresponding to pangolin fecal samples DG14 and DG18 were deposited in BioProject PRJNA641544, registered on 24/6/2020 by the Guangdong Institute of Applied Biological Resources (GIABR), and were reported as being sequenced on an Illumina Nextseq 550 instrument. None of the other positive samples reported in the paper (DG13, GZ1-2 and GZ5-2) have NGS datasets available. Sample DG14 was described as "Pangolin coronavirus from feces", and as such we expected a significant amount of PCoV reads to be found in the dataset. Instead, viruses of subfamily *Densovirinae* and the order *Caudovirales* were the most abundant, followed by Murine Leukemia Virus at 2% of viruses identified using STAT (Supp. Fig. 1). No PCoVs or SARSr-CoVs were detected using either STAT or fastv in either sample DG14 or DG18.

The DG14 and DG18 datasets were aligned to a set of SARSr-CoVs to determine if PCoV could be detected in the samples. Dataset DG14 contained seven reads (3 in the S gene, and 4 in the N gene) with highest homology to SARS-CoV-2 Wuhan-Hu-1, four with an exact match, and three with a 1nt SNV (Fig. 1). However 15 reads spanned an 81nt section of the genome covering the 3' end of the N gene and the non-coding region between the N gene and ORF10 (Figs. 2, 3). These reads, when analyzed using blastn, had the highest similarity to either PCoV GZ5-2 isolate DG14-3 or MP789. We consider two scenarios as plausible: 1) a mixture of SARS-CoV-2 and DG14-3 sequences, DG14-3 matching reads only located in a short 81nt region, which coincides with the PCR primers targeting fragments of the N and ORF10 genes (Li et al. 2021); 2) A novel PCoV, with higher identity to SARS-CoV-2 in the S and part of the N gene, which would indicate a third clade of PCoV in addition to the GD CoV and GX PCoV clades. We consider the hypothesis that the sequences stem from SARS-CoV-2 contamination and PCR-generated fragments more likely.

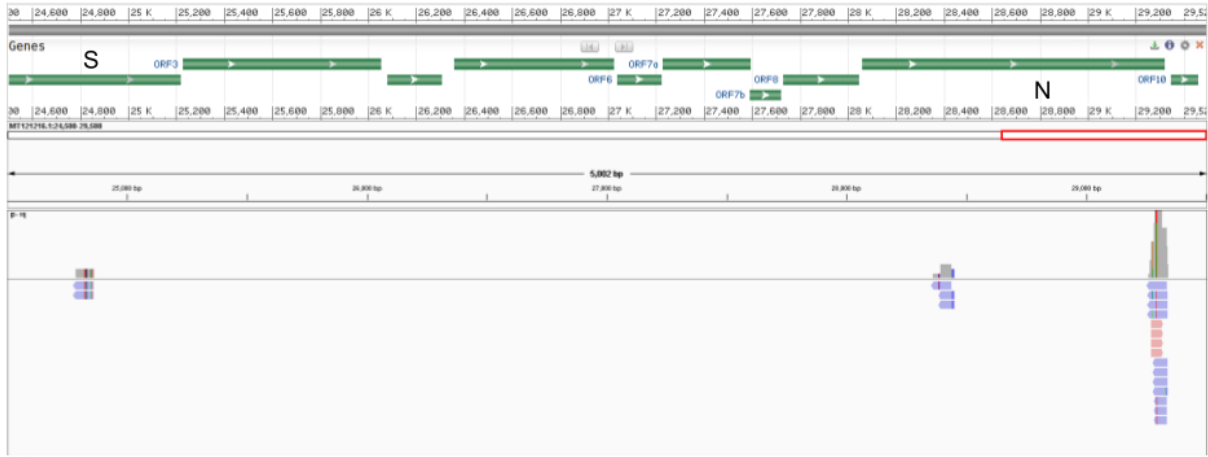

Fig. 1. Pooled fecal datasets DG14 and DG18 were aligned to PCoV MP789 using minimap2 and the 20 reads mapping to the genome plotted in IGV. Gene locations were plotted using NCBI MSA viewer.

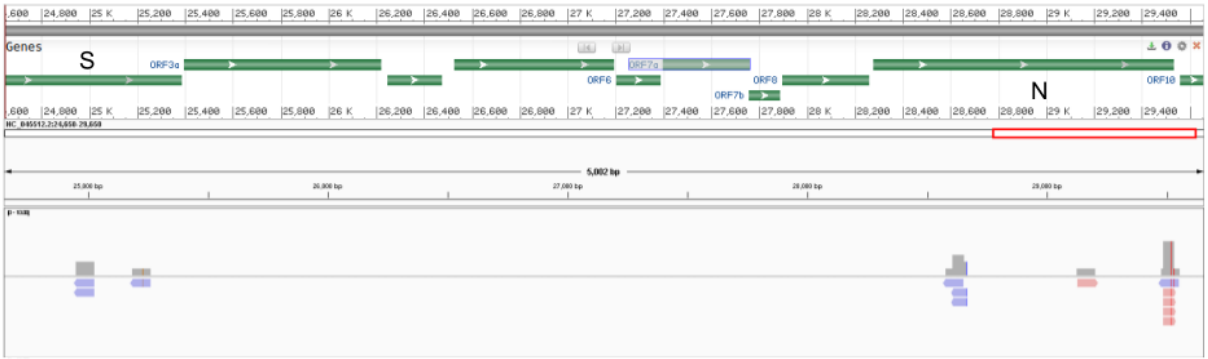

Fig. 2. Pooled fecal samples DG14 and DG18 were aligned to SARS-CoV-2 using minimap2 and the 12 reads mapping to the genome plotted in IGV. Gene locations were plotted using NCBI MSA viewer.

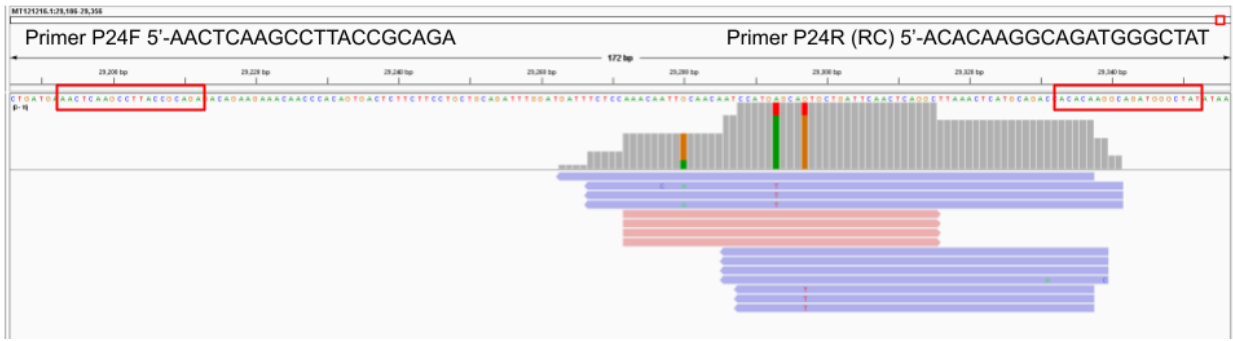

Fig. 3. 15 reads mapping to the 3' end of the N gene of PCoV MP789 are located in the specific primer capture region targeted by Li et al. (2021). Primer sequences highlighted in red boxes.

In sample DG14 bacteria comprised 98% of cellular organisms. Unexpectedly, however, *Hominoidea* and *Sus scrofa* comprised 18% and 12% of Eukaryota respectively (Fig. 4).

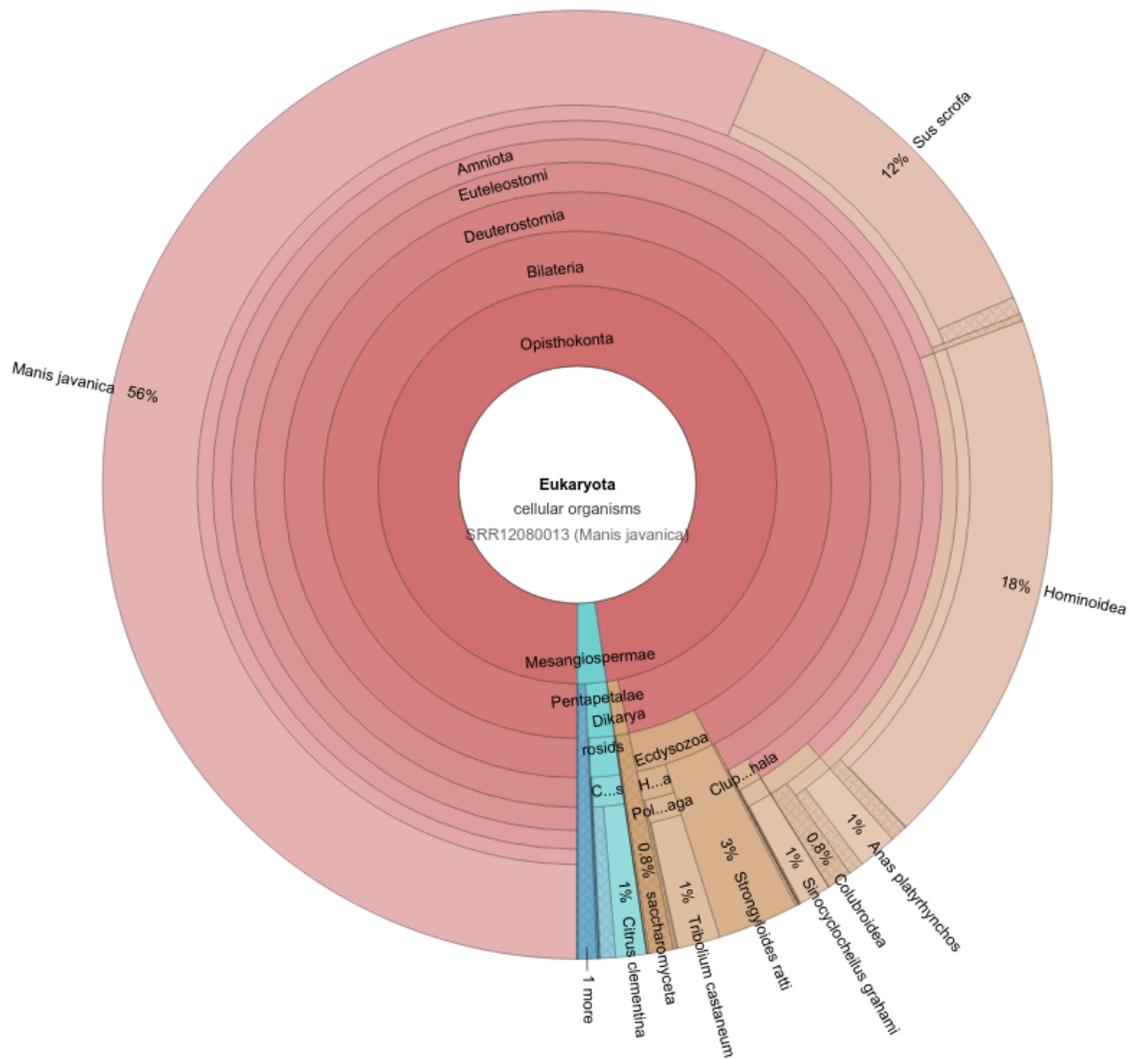

Fig. 4. NCBI STAT Krona analysis of pangolin feces sample DG14.

In sample DG18 bacteria comprised 58% of cellular organisms, with *Homo sapiens* comprising 0.2% of Eukaryota and *Manis javanica* comprising the remainder of the Eukaryota classification. We identified one 75nt read exactly matching SARS-CoV-2 Wuhan-Hu-1 in the N gene region.

The two datasets were then mapped to the entire NCBI mitochondrial database (Supp. Info. 2, Fig. 5). The MJ mitochondrial genome has the highest coverage out of all the mitochondrial genomes examined (99.8 % for DG14 and 99.0 % for DG18), consistent with the considerations above. The levels and range of contaminating sequences are remarkable. *Homo sapiens* shows 67.5 % (DG14) and 80.7 % (DG18) mitochondrial genome coverage. *Tenebrio molitor* (mealworm), shows 97.3 % (DG14) and 59.9 % (DG18) mitochondrial genome coverage.

Mealworm is used to feed captive pangolins (Zhang F. et al. 2021), and is consistent with prolonged captivity and interaction with humans. *Sus scrofa* shows 7.1 % (DG14) and 3.6 % (DG18) mitochondrial genome coverage, and is consistent with the presence of an expression plasmid contaminant containing the *Sus scrofa* CD163 ORF (described below).

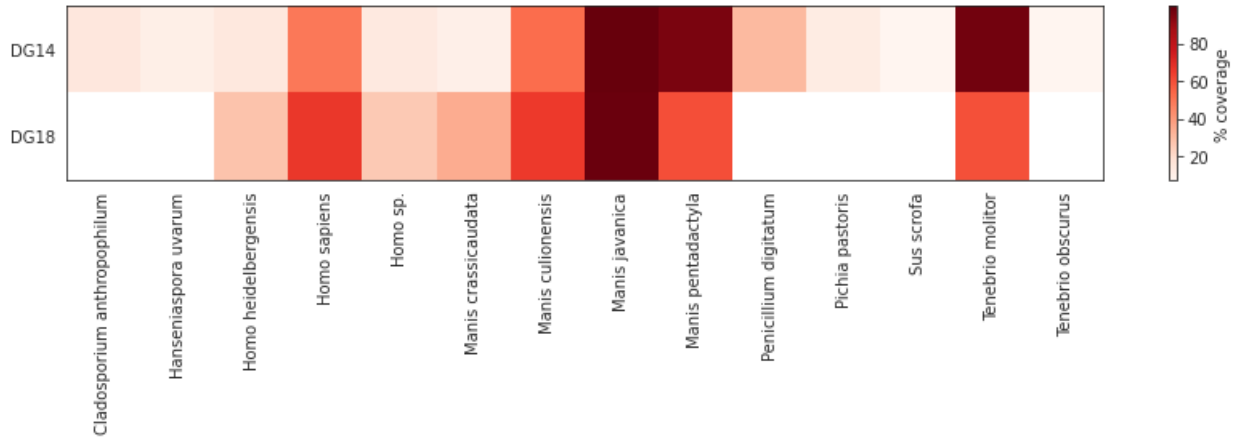

Fig. 5. Mitochondrial genomes with >7% coverage in datasets DG14 and DG18. Aligned using bowtie2 using default settings. Plotted using matplotlib.

To validate the STAT and full NCBI mitochondrial database mapping results, datasets DG14 and DG18 were aligned to *Manis javanica* (MJ), *Manis pentadactyla* (MP), *Homo sapiens* and *Sus scrofa* mitochondrial genes using 100% identity (Supp. Tables 1, 2). *Manis javanica* to *Manis pentadactyla* mitochondria ratios were 1.4:1 and 0.87:1 in DG14 and DG18 respectively.

The identity between the MJ and MP genomes is 99.35% with 107 SNVs and 6 gaps distributed over the genomes (Supp. Fig. 2). Reads in sample DG18 aligning to the MP mitochondrial genome with 100% identity were then re-aligned to the MJ mitochondrial genome using minimap2 (Supp. Fig. 3). 51 unanimous SNVs at positions where SNVs between MJ and MP mitochondrial genomes occur were counted. No other SNVs were found. The relative MJ/MP mitochondrial ratios in DG14 and DG18 and SNVs in 100% MP aligned reads in DG18 relative to the MJ mitochondrial genome indicating the potential presence of MP origin fecal material in the samples (Supp. Info. 1).

The DG14 and DG18 datasets were then pooled and *de novo* assembled using MEGAHIT. A python script was used to identify contigs with standard synthetic primer and promoter sequences. We identified a 7343 bp plasmid and used NCBI BLASTX to analyze the open reading frame (ORF) coding regions. Both DG14 (1586 reads) and DG18 (486 reads) had reads mapping to the plasmid. The largest ORF was found to have highest similarity to the full length CD163 protein (*Sus scrofa*) (1114/1115bp, 99.91% identity) (ADM07458.1) (Fig. 6). In the recovered plasmid, the CD163 gene is under control of the CMV promoter, which is used for high level expression in mammalian cells.

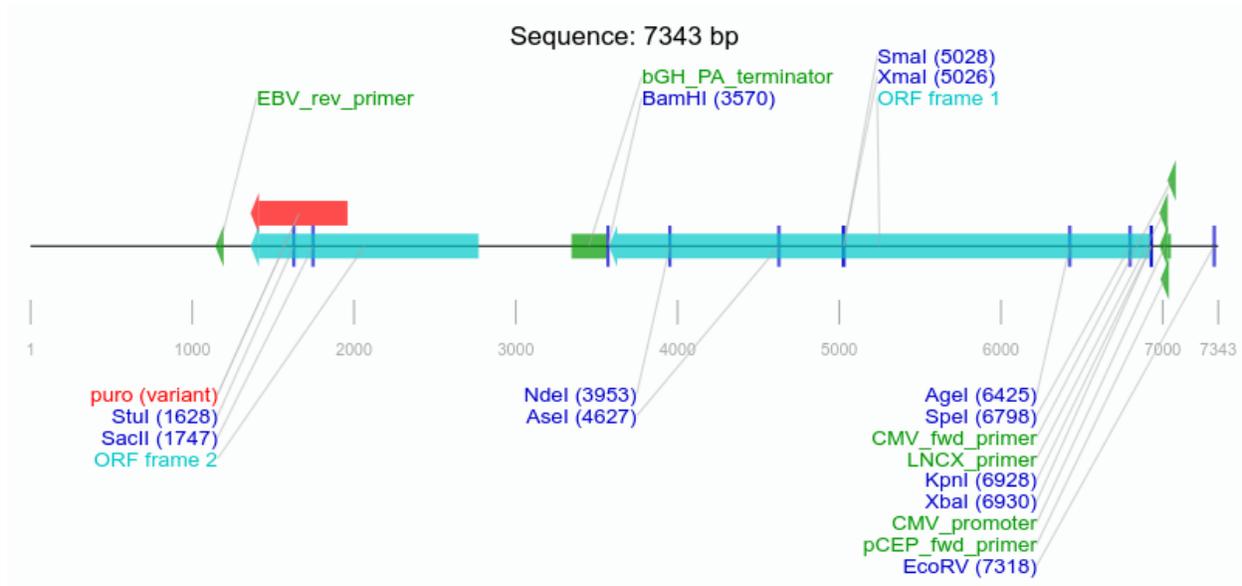

Fig. 6. Plasmid sequence identified in contig k79_145004 after *de novo* assembly of pooled DG14 and DG18 samples.

CD163 is a hemoglobin-haptoglobin (HbHp) complex scavenger receptor dominantly expressed by monocytes and macrophages (Kristiansen et al. 2001). Expression of CD163 by monocyte/macrophage cells is regulated by pro-inflammatory and anti-inflammatory mediators (Buechler et al. 2000). CD163 further functions as a receptor for bacterial and virus pathogens, an erythroblast adhesion receptor, and a receptor for tumor necrosis factor-like weak inducer of apoptosis (TWEAK) (Van Gorp. et al. 2010). Given the identification of *Sus scrofa* CD163 it is interesting to note that downregulation of expression of CD163 can restrict Porcine Reproductive and Respiratory Syndrome virus (PSSRV) infection (Zhu et al. 2020). Another widely researched pig-hosted virus with the potential significant detrimental effect on the pig industry is African Swine Fever Virus (ASFV), where CD163 may be involved but is not essential (Lithgow et al. 2014). However no PRRSV or ASFV sequences were identified in the datasets. Human CD163 expressing monocytes/macrophages may be an important component of immunopathology associated to SARS-Cov-2 infection (Gómez-Rial et al. 2020) and pig CD163 could potentially be used for analysis of SARSr-CoV macrophage response via immunofluorescence analysis. We note PCoV GD (EPI_ISL_410721) and PCoV-GX_P5L (EPI_ISL_410540) pseudotyped viruses have been demonstrated to infect porcine cells (Nie et al. 2021), with both PCoVs exhibiting had significantly higher infectivity than SARS- CoV-2 in porcine cells (ST and PK15).

A 5124 bp synthetic vector sequence was also identified containing the pACYC-F p15A origin (forward primer) and CAT-R, the 5' end of chloramphenicol resistance gene (reverse primer)

(Supp. Fig. 4) . BLASTX analysis of the largest ORF shows 100% identity (874bp) to T3/T7-like RNA polymerase [Salmonella phage SP6] (NP_853568.1). SP6 RNA polymerase has been used for *in vitro* RNA synthesis (Kreig and Melton, 1987; Stump and Hall, 1993), but the purpose of the vector found in this dataset is unclear.

A comparison of the number of reads mapping to PCoV MP789, SARS-CoV-2 Wuhan-Hu-1 and the two *de novo* assembled contigs containing plasmid sequences shows a c. 100 to 2000X higher number of reads mapping to synthetic plasmids than SARS2r-CoVs in these fecal samples (Table 1).

| Sample | MP789 | SARS-CoV-2 | 7343 bp plasmid (contig k79_145004) | 5124 bp plasmid (contig k79_60474) |
|--------|-------|------------|-------------------------------------|------------------------------------|
| DG14 | 15 | 6 | 1614 | 6636 |
| DG18 | | 1 | 491 | 2357 |

Table 1. Number of reads mapping to selected SARS2r-CoVs and contigs in fecal samples DG14 and DG18. Mapped using minimap2

The sequences for the N gene and 33nt of the non-coding region (NCR) between N and ORF10 for several SARSr and SARS2r-CoVs were aligned using MUSCLE in UGENE. The sequence lengths were 1306nt after multiple genome alignment. A maximum likelihood tree was then generated using MEGA11 with a GTR+G+I model. The GZ5-2 partial N gene and 33nt N to ORF10 NCR sequences published by Li et al. (2021) were 160nt in length and as such only comprise 12% of the N and partial NCR aligned region. The GZ5-2 isolates cluster with the Guangdong PCoV clade which forms a basal sister relationship to the SARS2r-CoV clade (Fig. 7). GX_WIV (Jones et al. 2022b), PCoV GX-5E and PCoV GX-P3B are positioned along a slightly more basal branch than other Guangxi PCoVs in the N and partial NCR region.

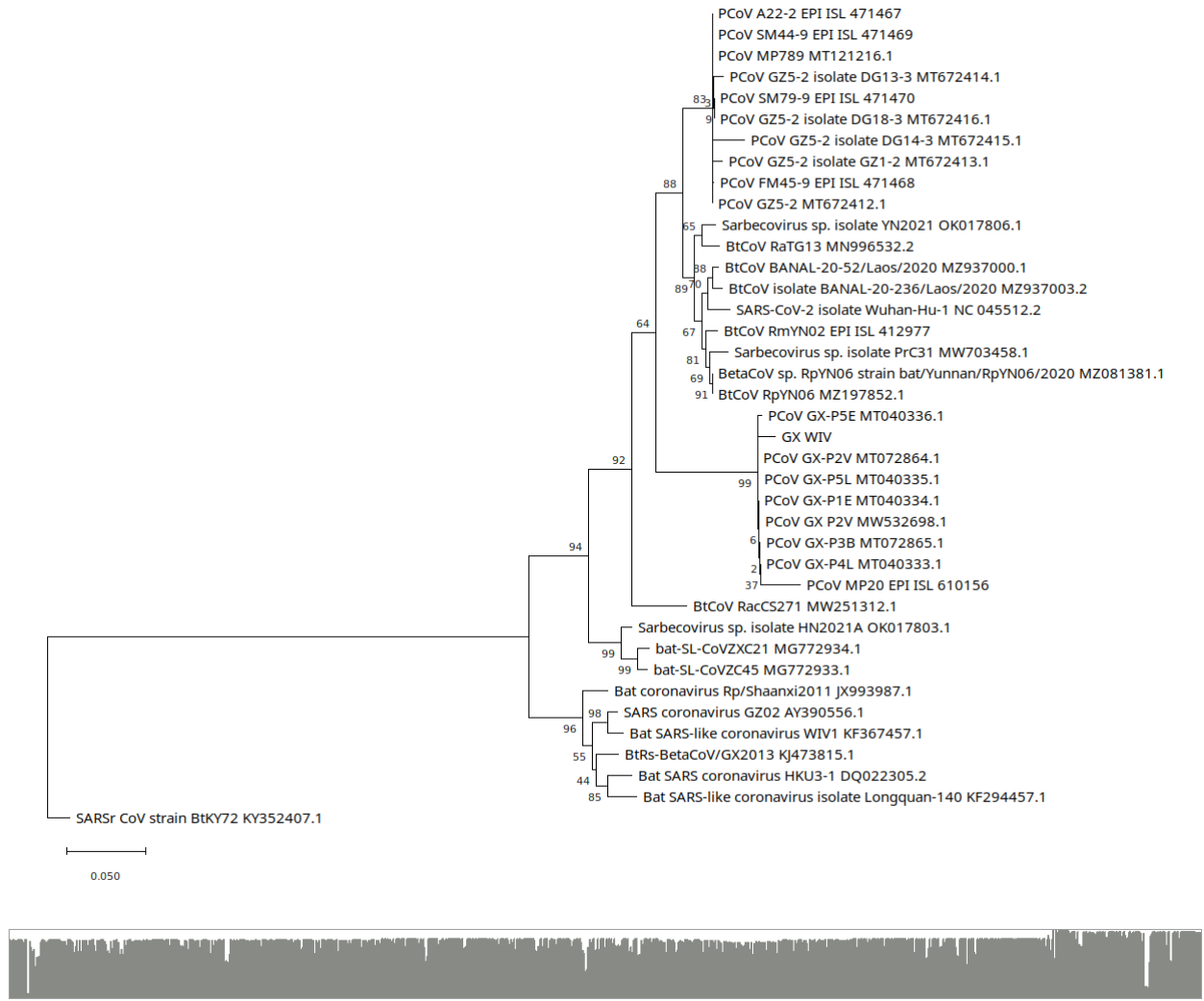

Fig. 7. Maximum likelihood tree for the N gene and partial non coding region (NCR) for selected CoVs. Generated using a GTR+G model with 1000 bootstrap replicates in MEGA11. All GZ5-2 PCoV isolates had only partial coverage of the N gene and NCR. The image at the bottom of the tree shows overall sequence similarity with position along the N gene.

A maximum likelihood tree was also generated using the short 160nt isolates of PCoV GZ5-2: DG14-3, DG18-3, DG13-3 and GZ1-2, with selected SARS2r-CoVs. A K2+G was found to have the lowest BIC score using model testing in MEGA11 (Supp. Fig. 5). The relationship of PCoV GZ5-2 isolates to GD PCoVs is unchanged.

We also note that relative to GD PCoVs and other SARSr-CoVs, GX PCoVs have a 6nt deletion near the 5' end of the N gene (Supp. Fig. 6). While both GD and GX PCoVs have 6nt and 2nt deletions relative to selected SARSr-CoVs, also in the N gene (Supp. Fig. 7).

*GX/P3B*

Three RNA-Seq SRA datasets were provided by Lam et al. (2020), two in support of Guangxi PCoVs, GX/P3B and GX/P2V, while a third dataset GD/P2S had all non-CoV reads filtered out from the dataset. The four other datasets provided were PCR Amplicon sequencing datasets with extreme enrichment of PCoV content (Table 2). Datasets GX/P3B and GX/P2V were sequenced on an Ion Torrent S5 XL platform. While read lengths in GX/P3B follow a log normal distribution (average of 52nt post fastp filtering), read lengths in GX/P2V follow a normal distribution (Supp. Fig. 8). Given the two samples were sequenced on the same platform, the reason for the skewed distribution of read lengths in GX/P3B is unknown. The short read lengths and abundant indels in GX/P3B make raw read alignment at a lower confidence than GD PCoV datasets analyzed by Jones et al. (2022a) (Supp. Fig. 9). It is interesting to note that Jones et al. (2022a) identified SARS-CoV-2 contamination of the GD/P2S and GX/P2V datasets which was identified using STAT analysis (Table 2). This indicates the usefulness of STAT analysis for contamination identification (Katz et al. 2021).

| Sample | Type | Strategy | Manis Javanica | Euarchontoglires | | | | Euteleosteomorpha | PCoV | SARSr/2r-CoV | Bacteria | Unidentified |
|---|---|---|---|---|---|---|---|---|---|---|---|---|
| | | | | | Simiiformes | Murinae | Chlorocebus | Salmo salar | | | | |
| GD/P2S | scale | Viral RNA | | | | | | | 57% | 6% | 0% | 37% |
| GX/P1E | intestine | Amplicon | 2% | | | | | | 96% | | 0.2% | 2% |
| GX/P5L | lung | Amplicon | 16% | | | | | | 81% | | 0.1% | 3% |
| GX/P5E | intestine | Amplicon | 0.6% | Trace | | | | | 98% | | 0.07% | 2% |
| GX/P4L | lung | Amplicon | 53% | | | | | | 36% | | 0.2% | 11% |
| GX/P3B | blood | Viral RNA | 9% | 0.7% | 0.6% | 0.02% | | | 0.007% | | 0.08% | 90% |
| GX/P2V | intestine-lung | Viral RNA | | | | | 80% | 2% | 1% | 0.03% | 4% | 10% |

Table 2. NCBI STAT analysis of datasets in BioProject PRJNA606875.

After *de novo* assembly using MEGAHIT, contigs from GX/P3B were aligned to a combined NCBI univec and viral database, and contigs matching viral and plasmid sequences were then analyzed using blast against a local copy of the nt database. 16 contigs matching GX-PoVs were found, as well as 4 contigs of lengths from 106 to 662nt matching with highest identity to various cloning vectors (Supp. Table 3). In addition to synthetic vectors in the dataset, a further contamination concern is that 4.3% of 100% identity mitochondrial matches were to the *Homo sapiens* mitochondrial genome (Jones et al. 2022a).

We additionally aligned reads in GX/P3B to all mitochondrial genomes in the NCBI database and apparently detected a wide range of contaminating mitochondrial sequences (Supp. Info. 3). However, we determined that this is due to the presence of a substantial number of reads of less than 22 nt in the dataset (Supp. Fig. 10), which was sequenced on an Ion Torrent S5 XL, and has an average read length of 37nt prior to and 52nt post fastp processing (Chen et al. 2018). Consequently, while mapping to all mitochondrial genomes in the NCBI is a powerful approach for detecting contaminating sequences, short sequences should be removed from the dataset first.

We then aligned all reads longer than 32nt in GX/P3B to the Homo sapiens genome assembly GRCh38.p13 with 100% identity then filtered out other animal genomic matches as described in methods. We found 28562 reads matching the human genome. We then filtered out reads shorter than 80nt leaving 163 reads which were analyzed using local blast against the nt database, confirming human or primate DNA/RNA sequence matches outside of the mitochondrial genome (Supp. Info. 3).

Analysis of bacterial taxonomy using kraken2 indicates contamination of blood sample GX/P3B by plant-root and soil hosted Alphaproteobacteria of the *Sphingomonadales*, *Hyphomicrobiales*, *Rhodobacterales*, and *Caulobacterales* orders (Supp. Figs. 11, 12).

In the Terrabacteria group several obviously contamination-related bacteria were identified. In the Firmicutes phylum, *Selenomonas ruminantium* subsp. *lactilytica*, a sheep rumen hosted bacterium (Kaneko et al. 2015) was found to comprise 1% of total bacteria (Supp. Fig. 13). While in the Actinobacteria phylum, *Streptomyces*, dominantly found in decaying vegetation comprised 1% of bacteria. Furthermore wastewater hosted *Microlunatus phosphovorus* NM-1 (Nakamura et al. 1995) comprised 0.4% of total bacteria.

While Proteobacteria and Terrabacteria group total percentages of bacteria in sample GX/P2V a virus isolate hosted in Vero-E6 cells, is similar to GX/P3B, the phyla differ significantly. In GX/P2V Proteobacteria are dominated by Gammaproteobacteria with *Halomonas* sp. JS92-SW72 comprising 94% of the phylum and 25% of total bacteria (Supp. Fig. 14). This species was previously found to comprise approximately 4-7% of microbiota in human lung BALF samples from patients diagnosed with pneumonia (Hong et al. 2021). As a human lung hosted pathogen we infer the presence of this species as a significant fraction of total bacteria in this Vero-E6 cell culture sample to be contamination-related. In the Betaproteobacteria, domestic sewage soil hosted bacteria *Hydrogenophaga* sp. NH-16 (SAMN08998297) comprised 4% of total bacteria. In GX/P3B this species comprised 0.3% of bacteria potentially indicating a similar source for this bacteria in the two samples.

In the Terrabacteria group, *Tenericutes* was the dominant phylum in GX/P2V with mycoplasma comprising 17% of bacteria. The species *Mycoplasma hyorhinis* GD-1 comprised 10% of total bacteria. This finding appears to be at odds with the statement by Lam et al. (2020) that all cells were "free of mycoplasma contamination". While in the *Firmicutes* phylum, *Staphylococcus aureus* comprised 1% of bacteria.

***Liu et. al. (2019)***

Jones et al. (2022a) aligned each SRA in BioProject PRJNA573298, labeled as 'Viral diversity and pathogens of dead Malay pangolin samples',  to a set of mitochondrial genomes using 100% matches. Lung04, Lung03, Lung08 and Lung02 were found to have the highest *Homo sapiens* mitochondrial genome matches as a percentage of all mitochondrial matches. Here we aligned each SRA dataset in this BioProject to all mitochondrial genomes on NCBI as of 22/05/2022 (Supp. Info. 4.2). *Homo sapiens* mitochondrion at >10% genome coverage is evident in all samples except LymphA01 (5.9% coverage) and *Mus musculus* mitochondrial genome contamination at >10% coverage is found in 57% of samples. Interestingly *Panthera tigris* (tiger) mitochondrial sequences are only present in samples with no *Mus musculus* mitochondrial content (Fig. 8; Supp. Fig. 15). While *Mus musculus* (and *Homo sapiens*) mitochondria were found in samples sequenced on two different instruments on 4 separate runs, all tiger mitochondrial reads were found in samples sequenced using machine A00192 flowcell id HHYMLDSXX in run number 246. Of the four samples containing tiger mitochondrial reads, three were sequenced on lane 1 and one on lane 2 of the same run (Supp. Info. 4.1).

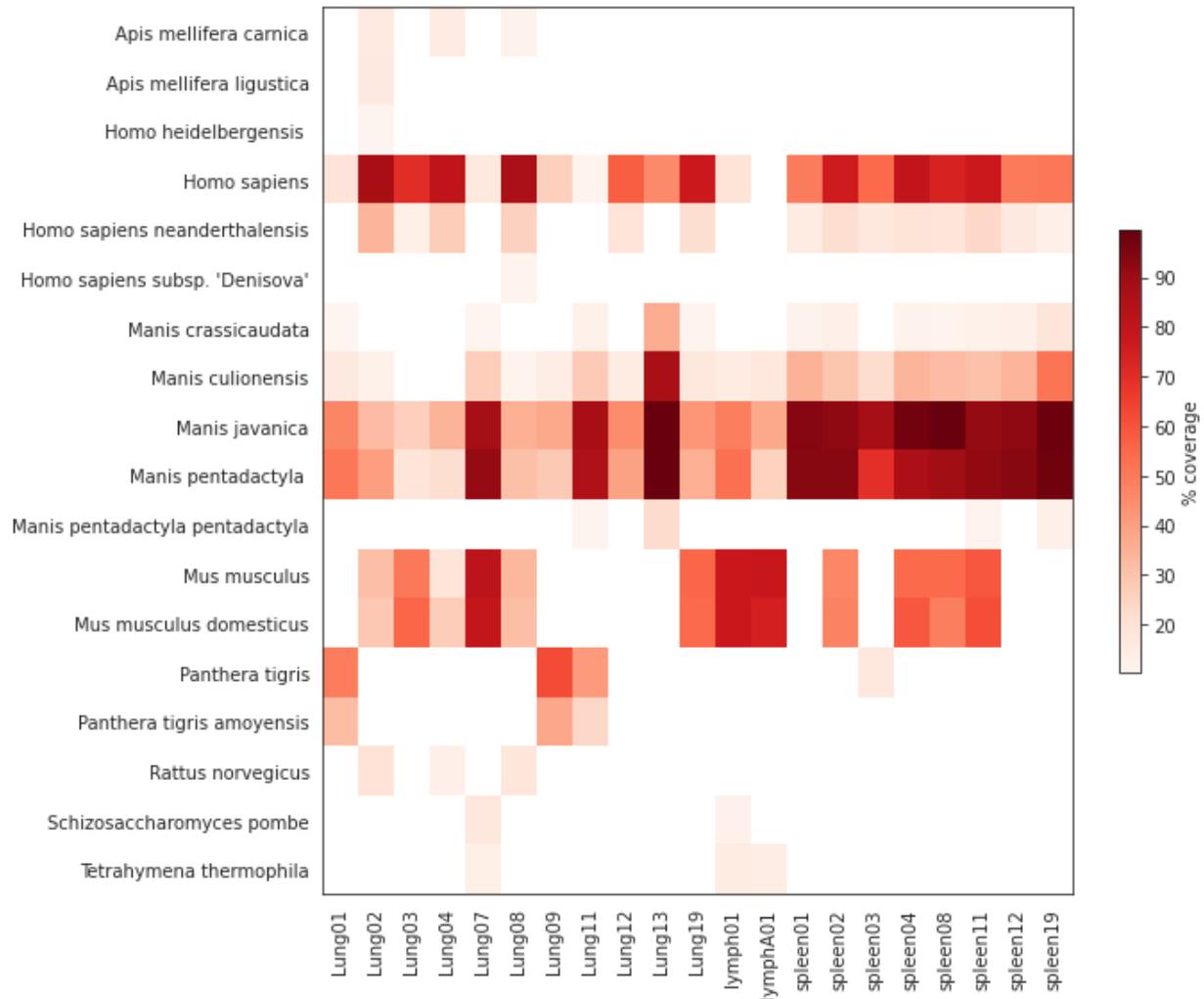

Fig. 8. Mitochondrial genome coverage >10% for each sample in BioProject PRJNA573298.

Taxonomic analysis of each dataset in BioProject PRJNA573298 was run using Kraken2 against the standard nt database and the bacterial taxonomy was reviewed. While the Terrabacteria Group and Proteobacteria were most abundant, the distribution and range of bacteria was quite varied across samples. Interestingly, *Staphylococcaceae* spp. were found at high levels in samples Lung09, Lung08, Lung07, at 13%, 22%, and 8% of bacterial content (Supp. Figs. 16-18). Plant root inhabiting bacteria of the *Hyphomicrobiales* including *Hyphomicrobiaceae*, *Xanthobacteraceae* and *Rhizobiaceae* were found in most samples with family level percentages of total bacterial content generally 0.01% to 3% however *Xanthobacteraceae* comprised 9% of bacteria in Lung09 and *pseudolabrys sp. FHR47* a species of the family *Xanthobacteraceae* comprised 58% of bacteria in spleen12 (Supp. Fig. 19). A genus also inhabiting plant root systems, *Sphingomonas* was found at highest levels in Lung13 where it comprised 25% of bacteria (Supp. Fig. 20). *Mycoplasma* spp. were identified at 7% of bacterial content in Lung13 and Lung12 (Supp. Fig. 21). *Selenomonas ruminantium* subsp. *lactilytica TAM6421* a bacteria isolated from sheep rumen (Kaneko et al. 2015) was identified in samples spleen11, spleen08,

spleen04, spleen03, spleen02 and lung12 at 1%-8% of total bacteria content. Curiously the same sheep rumen hosted bacteria *Selenomonas ruminantium* subsp. *lactilytica* was identified in Guangxi pangolin CoV dataset GX/P3B.

## Discussion

Previously we identified multiple issues with pangolin metagenomic datasets supporting pangolin CoVs and raised concerns for attribution to a definitive host (Jones et al. 2022a). The Liu et al. (2019) datasets, which were used to generate the Guangdong (GD) PCoV MP789 and PCoV GD_1 genomes  are consistent with SARS2r-CoV sequences being a result of inadvertent contamination and misattribution to a pangolin host rather than genuine infection. This supposition is supported by the following observations:

1) 5 of the 6 SRA datasets in Liu et al. (2019) that contained SARS2r-CoV sequences were found in pangolin major organ samples with significant bacterial contamination, the sixth contained only 4 reads (Fig. 9). An association of anomalously high bacterial levels and virus contamination in RNA-Seq datasets has been noted previously, where a human BALF sample with bacteria content at an anomalously high level was most heavily contaminated by synthetic plasmids containing the Influenza HA gene sequence, and Nipah virus sequences within synthetic vectors Quay et al. (2021a, 2021b).

2) Human and mouse genomic origin contamination of the datasets is significant, with multiple independent taxonomy classification methods indicating *Homo sapiens* genomic origin content lower bound of approximately one percent in any sample and in most cases several percent (Figs. 10, 11).

3) After pooling all SRAs in the BioProject, *Homo sapiens* mitochondrial coverage was found to be 93.9%, with all SRAs containing reads with 100% identity matches to the *Homo sapiens* mitochondrial genome (Fig. 12).

4) Non pangolin hosted viruses contained in the samples were found with abundance levels on the order of PCoV MP789/GD_1 read content.

5) PCoV MP789/GD_1 binding to pangolin ACE2 is tenfold weaker than binding to human ACE2 (Wrobel et al. 2021) which is unexpected for a pangolin-hosted virus. This result, however, is consistent with GD PCoV being a human adapted CoV contamination of the Liu et al. (2019) datasets (Jones et al. 2022a).

6) Liu et al. (2019) identified a novel Sendai virus (SeV) strain in 6 of 11 pangolins sampled and note this was the most widely distributed pathogen in the samples. The authors misattributed the closest strain they identified Ohita-M1 (AB005795.1) as a human hosted SeV and strongly suggested the possibility that the SeV is transmitted between pangolins and humans. However the Ohita-M1 strain was isolated from a viral outbreak in an animal laboratory by passaging the virus in mice (Itoh et al. 1997). The SeV in the Liu et al. datasets had by far the highest read count of any virus found (Jones et al. 2022a). Yang et al. (2021) isolated the strain and identified the novel SeV as pangolin-hosted strain M5. It seems highly unlikely that a single batch of pangolins rescued by the Guangdong Wildlife Rescue Center (Chan and Zhan, 2020) would have contracted two novel viruses that jumped from other species, both viruses being extremely rare, with pangolin SeV M5 not yet detected in any other pangolins sampled outside of those from the single batch at the Guangdong Wildlife Rescue Center and PCoV MP789/GD_1/GD_P2S only identified in a small number of pangolins (Xiao et al. 2020; Liu et al. 2020, Lam et al. 2020).

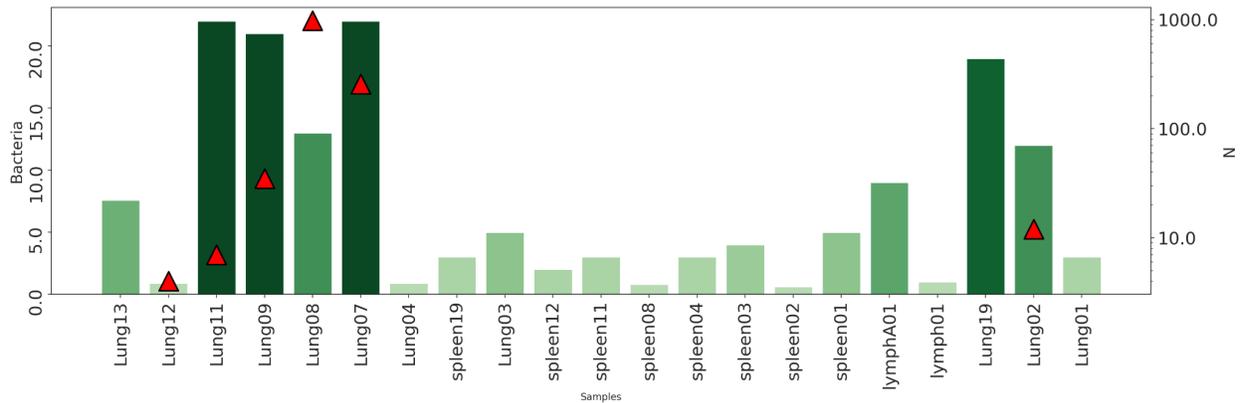

Fig. 9. Bacterial percentages of pangolin major internal organ SRA datasets sequenced by Liu (barplot, left linear scale) and number of pangolin CoV reads (red triangles, right log scale). For all but Lung12 which contained only 4 reads of Pangolin CoV MP789, PCoV MP789 is found in samples with bacteria comprising >=12% of the RNASeq dataset. See Jones et al. (2022) for full method description.

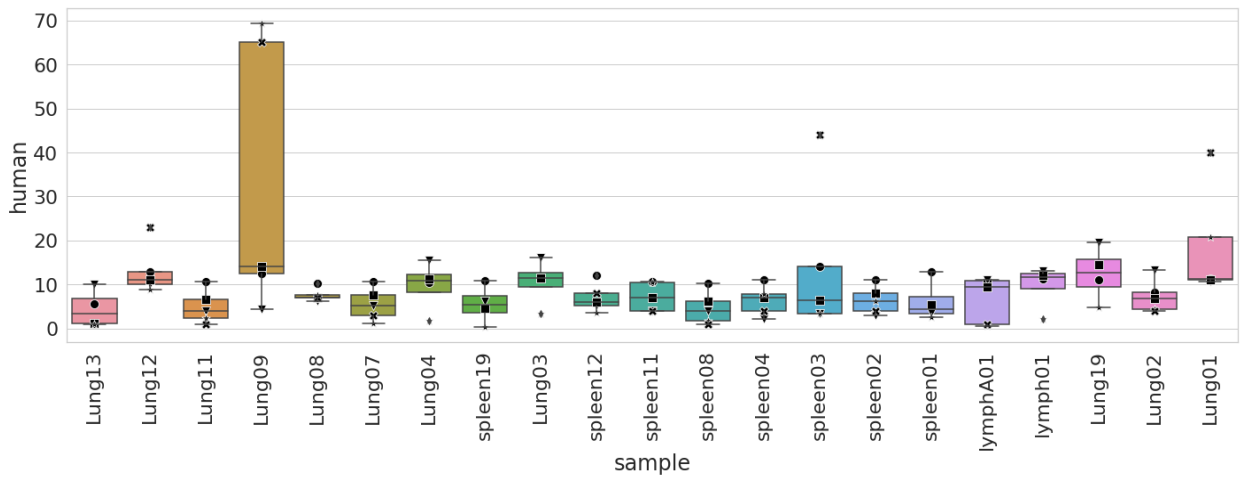

Fig. 10. Box plot of *Homo sapiens* genome matching reads as a percentage of total reads matching any of the full genomes for the four species: *Manis javanica, Manis pentadactyla, Homo sapiens* and *Mus musculus*. Five methods were used for estimation: SerialAlign using 'very sensitive' setting (circle), Disambiguate (triangle), SerialAlign using bowtie2 with the "--local --score-min L0,0" parameter (square), Seal using the 'toss' parameter (star), and NCBI STAT (x). STAT percentages are plotted for Simiiformes or lower taxonomic rank. For Lung04, Lung03, Spleen01, Lymph01 and Lung19 28%, 39%, 4%, 40% and 23% Euarchontoglines respectively were classified by STAT but not sub-classified and not plotted here. Box extends from the lower to upper quartile values, a horizontal line delineates the median. Whiskers extend to 1.5X of the interquartile range of the data. Outliers are drawn as diamonds. See Jones et al. (2022) for full method description.

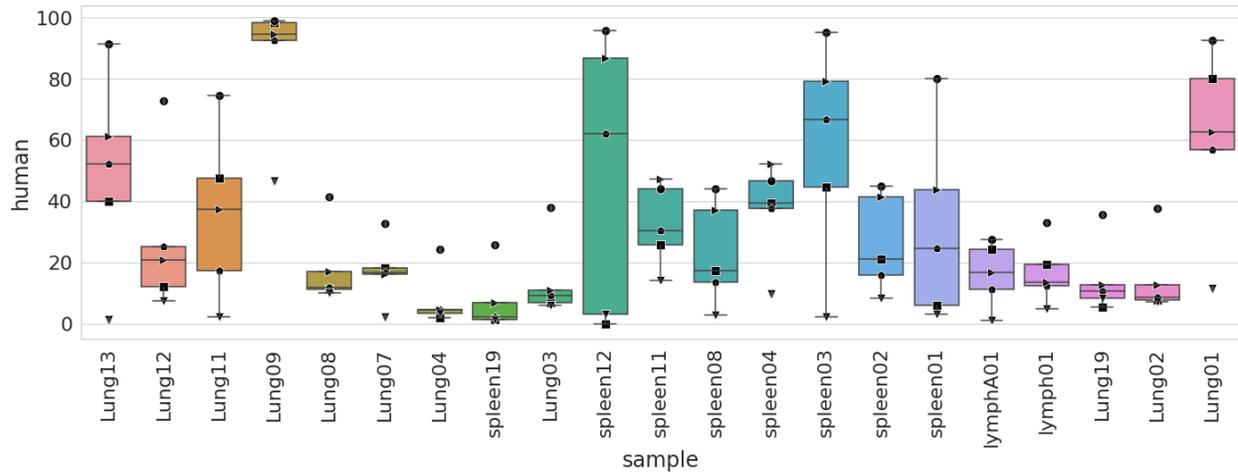

Fig. 11. Box plot of percentage of *de novo* assembled contigs matching the *Homo sapiens* genome as a percentage of reads matching *Manis javanica, Manis pentadactyla, Homo sapiens* and *Mus musculus* genomes. Five methods were used for estimation: SerialAlign using bowtie2 with the "--local --score-min L0,0" parameter (square), SerialAlign using 'very sensitive' setting (circle), ConcatRef (pentagon), Blast (triangle right), and Disambiguate (triangle down). Although the longer lengths of contigs allows greater accuracy than read alignments, the absolute number of contigs may not reflect the true overall percentage content of matching genomic sequences in the SRAs. See Jones et al. (2022) for full method description.

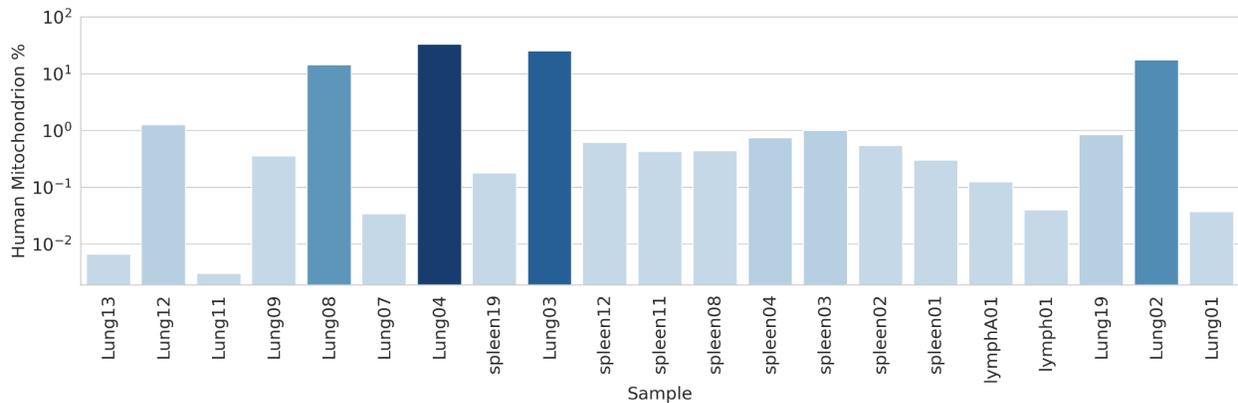

Fig. 12. Percentage of *Homo sapiens* mitochondrial genome matches for reads with 100% identity matches to selected mitochondrial genome sequences. Mitochondrial genome sequences identified were: *Manis javanica, Manis pentadactyla, Homo sapiens, Mus musculus* and *Panthera tigris* genomes. Percentage plotted in log scale, Lung08, Lung04, Lung03 and Lung02 has the highest human mitochondrial percentages at 14%, 42%, 25% and 18% of all mitochondrial genome matches respectively. See Jones et al. (2022) for full method description.

The PCoV GD_1 and PCoV MP789 genomes were derived from the Liu et al. (2019) pangolin datasets by Xiao et al. (2020) and Liu et al. (2020) using additional pangolin sequence datasets. However, a key dataset GD1 provided by Xiao et al. (2020), without which the PCoV GD_1 genome cannot be assembled, contains PCoV GD_1 genome sequences hosted in pEASY-tub/hptII cloning vectors (Jones et al. 2022a). Further issues with the BioProject are that pangolin lung datasets Z1 and M5 contain African Swine Fever Virus and Porcine Circovirus 2 sequence in plasmids (Jones et al. 2022a). In the M1 pangolin SRA dataset provided by Xiao et al. (2020), reads aligned to the GD_1 genome exhibit a circular read pattern of coverage, where the reverse complement of misaligned read ends across a 5 read wide zone matched the genome sequence at the other end of this region. We interpreted this pattern to be consistent with circularized cDNA generated during the ligation step of molecular cloning (Jones et al. 2022a).

Liu et al. (2021) state that an amplicon dataset (journal_ppat_1009664_s001) and RNA-Seq dataset GZ1-2, in addition to Lung07 and Lung08 from Liu et al. (2019), were used to generate the MP789 genome sequence. However we found a 3nt gap in coverage of the MP789 genome and SNVs for the two reads covering position 5731nt relative to the reference genome when the datasets provided by Liu et al. (2020) and Liu et al. (2021) were used, and question how the full genome could have been generated (Jones et al. 2022a).

Lam et al. (2020) sequenced a keratinous pangolin scale and derived the PCoV GD/P2S genome. Surprisingly, Lam et al. (2020) obtained 2.6 times more SARS2r-CoV reads from a pangolin scale than from any organ sample sequenced by Liu et al. (2019), and unusually, did so using two separate sequencing machines and combined the result. However we estimated 99.995% of reads had been filtered out, yet still contamination was detected with 10% of the reads matching the SARS-CoV-2 Wuhan-Hu-1 genome (Jones et al. 2022a). No RNA-Seq datasets are available for the GD related PCoV genomes FM45-9, SM44-9, SM79-9 published on GISAID by Xiao K.

The partial GD PCoV sequences found here in samples DG14 and DG18 do little to provide any support for GD PCoV epidemiological infection models (Li et al. 2021). Only 22 reads in total were recovered, with 7 reads having their closest match to SARS-CoV-2, indicating possible contamination by SARS-CoV-2 sequences during preparation or sequencing. The remaining 15 reads were located in a small region at the 3' end of the N gene wholly located in the PCR primer extraction region used by Li et al. (2021) for CoV detection, indicating contamination with PCR amplicons.

The review here of mitochondrial genome alignments for each RNA-Seq dataset in BioProject PRJNA573298 sequenced by Liu et al. (2019) adds some further detail to the analysis by Jones et al. (2022a). *Homo Sapiens* mitochondrial contamination was found in all samples while *Mus musculus* mitochondrial contamination was found in 13 of 21 samples, sequenced in 2 instruments, 4 runs and 4 separate flow cell ids. Tiger mitochondrial sequences were only found in 4 SRA's, all sequenced on the same instrument, flowcell id and run, but one of the 4 SRA's was sequenced on a separate lane. We infer the most likely contamination scenario is that human and mouse contamination occurred prior to sequencing, possibly due to a spill or contamination during a preparation stage. However tiger contamination may have occurred due to index hopping where reads from a project sequencing tiger tissue or cells were incorrectly assigned to *Manis javanica* during demultiplexing. A common cause for this is the presence of unligated adaptors in pooled samples prior to the sequencing.

The four cloning vector sequences present in contigs generated from *de novo* assembly of dataset GX/P3B could not be identified with certainty due to their relatively short lengths ranging from 106 to 662nt. The presence of cloning vectors, and the corroboration of the identification of *Homo sapiens* genomic origin sequences with the identification of *Homo sapiens* mitochondrial sequences by Jones et al. 2022a, raise concern over the reliability of the only non-enriched/non-heavily filtered pangolin tissue RNA-Seq dataset supporting assembly of a GX-CoV. It is quite extraordinary that there has been such a paucity of data provided to support the finding of GX CoVs in wild pangolins. This is even more extraordinary if one considers the importance of the GX CoV clade in supporting pangolins as the only other species to bats, harboring SARS2r-CoVs in the wild.

The bacterial contamination of GX/P3B has some similarities with bacterial contamination of the SRA datasets in BioProject PRJNA573298 sequenced for Liu et al. (2019) by the Guangdong Magigene Biotechnology Co. Ltd. (广东美格基因科技有限公司) in Guangzhou. The Guangxi PCoV datasets in BioProject606875 including GX/P3B were sequenced by Vision Medicals (微远基因, also known as Guangzhou Weiyuan Gene Technology) also in Guangzhou, Guangdong Province. Both dataset GX/P3B and datasets in BioProject PRJNA573298 had significant percentages of plant-root and soil-hosted bacteria and both contained the sheep rumen hosted species *Selenomonas ruminantium* subsp. *lactilytica*. One possibility to explain the occurence of a sheep rumen hosted bacteria in pangolin blood, spleen and lung samples is that sheep antisera could have contaminated the samples (Amer et al. 2018; Schmidt et al. 2018). The bacterial co-occurrences could be coincidental or potentially stem from reagents used during sample preparation. Another possibility is that as both datasets were sequenced in Guangzhou, the datasets were prepared and sequenced at the same sequencing center, with protocols and methods leading to endemic contamination that would be apparent even in samples sequenced months apart.

It is worth noting how unusual pangolins are as a proposed natural SARS2r-CoV reservoir (Zhang T. et al. 2020; Peng et al., 2021), intermediate SARS-CoV-2 host (Lam et al. 2020; Xiao et al, 2020) or incidental host (Lee et al., 2020, Li et al. 2021). Malayan pangolins (*Manis javanica*) and Chinese pangolins (*Manis pentadactyla*) are ground dwelling animals inhabiting grasslands and forests with diets of ants, termites and insect larvae (Hua et al. 2015). Pangolins are solitary except during an approximately 3 month mating season, where a male will mate with a female for 3-5 days (Hua et al. 2015). Although Giant pangolins (*Smutsia gigantea*) have been found to cohabit with Microchiroptera of the *Hipposideridae*, *Emballonuridae* and *Miniopterus* species in underground burrows in Gabon (Lehmann et al. 2020), no evidence has been found to show that *Manis javanica* or *Manis pentadactlya* in Asia co-inhabit burrows with bats.

Given that a SARS2r-CoV animal host would need to have a high population density to allow natural selection for virus adaptation (Andersen et al. 2020), pangolins are an unlikely natural or intermediate host. Indeed, in a study of 334 wild pangolins in Malaysia and between August 2009 and March 2019 no coronaviruses, Sendai virus or other potentially zoonotic viruses were detected (Lee et al. 2020), and the authors of the study infer that *Manis javanica* do not exhibit coronavirus shedding or endemic coronavirus infection at a population level.

In China between November 2019 and March 2020, ELISA antibody tests to SARS-CoV-2 in serum samples from 17 pangolins collected from undisclosed locations were all negative (Deng et al. 2020). No trace of SARSr-CoVs were found in 93 pangolin samples from BioProject PRJNA529540 sequenced by Hu et al. (2020) (Zhang D. 2020). Jones et al. (2022a) did not find SARS2r-CoV sequences in 7 RNA-Seq datasets in BioProject PRJNA610466 sequenced from Malayan pangolins collected from the Guangdong Provincial Wildlife Rescue Center by Li et al. 2020). Furthermore, no SARS2r-CoV matches were identified in 3 Guangxi pangolin datasets sequenced by Wenzhou University in BioProject PRJNA749865 (Jones at al. 2022). No SARS2r-CoV infection was detected in 21 Malayan pangolin or 12 Chinese pangolins of zoo and wild origin from Zhejiang province (SARSr-CoV sequences detected in the datasets were found to be sourced from contamination (Jones et al. 2022a) sampled by He et al. (2022). Finally, in a metagenomic study, Hu et al. (2020) undertook RNA-Seq of Malayan pangolins collected in China in 2017 but failed to detect any SARSr-CoVs.

An incidental host scenario invokes a temporarily higher population density than a natural host model, where pangolins are smuggled in batches, possibly in hundreds for rare large batches. The major destinations in China for illegally traded pangolins in the 2007-20016 period were Gungdong, Guangxi and Yunnan provinces and Hong Kong SAR (Xu et al. 2016). Main trafficking routes to Yunnan were from Vietnam and Myanmar overland, to Guangxi via fishing vessels from Vietnam and to Guangdong from Vietnam, Malaysia and Indonesia also via fishing boats (Xu et al. 2016). This model invokes spillover from bats or possibly another intermediate host, over to pangolins, and requires that bats (or other intermediate host) be smuggled with pangolins in close quarters to allow cross-infection. The model is problematic however, as the same type of bats carrying the specific strain of virus is required to repeatedly infect pangolins for the virus to be found in different pangolin batches. Furthermore, given the time constraint applied by smuggling duration of days to possibly months extremely limited natural evolution for pangolin host adaptation could occur, and the virus would need to be pre-adapted to be able to infect and transmit between pangolins.

A clear association of bat host family with betacoronavirus family is apparent, whereby Merbecoviruses (MERSr-CoV, HKU4r-CoV and HKU5r-CoVs) are strictly restricted to the *Vespertilionidae* family (Latinne et al. 2020), and Sarbecoviruses (HKU3r-CoV and SARSr-CoVs) CoVs originated in *Rhinolophidae* bats and are by far dominated by *Rhinolophidae* hosts (Latinne et al. 2020). However, at a species level bat CoVs are restricted more by geography than by bat species (Yu et al. 2019; Liang et al. 2021), with Sarbecoviruses tending to phylogenetically cluster by geographic region (Hu et al. 2017; Yu et al. 2019). Given the geographical restriction on Sarbecovirus diversity, the discovery of GX_ZC45 by Jones et al. (2022c), likely sourced from bats in Zhoushan city, Zhejiang Province raises questions as to the provenance of the GX-related PCoVs.

## Conclusion

We have found that two fecal samples DG14 and DG18 sequenced by Li et al. (2021) provided in support of GD PCoV infection of pangolins in Guangdong contain CoV read distributions consistent with PCR amplicon contamination. Multiple reads aligning to GD PCoV MP789 are located within the PCR capture region of the N gene, and nowhere else in the genome. The few other reads outside the PCR capture region best align to SARS-CoV-2 rather than GD PCoV MP789 or GX PCoVs. However we cannot rule out that a third lineage of PCoVs exists, with closer homology to SARS-CoV-2 in parts of the S and N genes than GD or GX PCoVs. We analyzed *de novo* assembled contigs from the Guangxi PCoV dataset, GX/P3B, and identified synthetic vectors in the dataset with far higher read mapping counts than SARSr-CoV genomes. Also in sample GX/P3B, we confirm the presence of human genomic origin sequences and infer bacterial contamination of the sample based on bacterial taxonomy. We aligned all pangolin organ datasets sequenced by Liu et al. (2019) to all mitochondrial sequences on NCBI and interpret that human and mouse contamination stem from upstream contamination prior to sequencing, and infer that tiger genomic origin material may be the result of index hopping and demultiplexing errors when tiger tissue or cells were sequenced on the same sequencing run. The DG14 and DG18 datasets add little to existing datasets supporting GD PCoV infection of pangolins which are widely contaminated and incomplete (Jones et al. 2022a). While GX PCoV RNA-Seq datasets are in most cases either heavily enriched or filtered, low quality, or not published. As such it is difficult to verify the attribution of either GD or GX PCoVs to a pangolin host. It is unfortunate that, given the importance of the GD and GX PCoV clades in identifying the origin of SARS-CoV-2, RNA-Seq datasets supporting the assembly of these viruses are not better sequenced and transparently shared.

## Methods

*DG14 and DG18*

Datasets DG14 and DG18 were aligned to 6 *Sus scrofa* mitochondrial genome sequences and *Sus scrofa* mitochondrion (MT199606.1) found to have overall highest coverage and read count. Samples DG14 and DG18 were aligned to each of the following *Sus scrofa* mitochondrial genomes using the bowtie2 v2.4.2 with the alignment option "--score-min L,0,0": MK688993.1, MK858173.1, MT199606.1, NC_000845.1, NC_012095.1 and KC469586.1.

Samples DG14 and DG18 were aligned to PCoV MP789 (MT121216.1) and SARS-CoV-2 (NC_045512.2) with polyA tail removed and contigs using minimap2 using the following parameters "-MD -c -eqx -x sr -2 -t 32 --sam-hit-only --secondary=no"

Contigs with synthetic primer/promoter sequences identified were analyzed using addgene sequence analyzer: https://www.addgene.org/analyze-sequence/

Multiple sequence alignments were generated using the MUSCLE algorithm (Edgar, 2004) in UGENE v42.0 (Okonechnikov et al. 2012).

Phylogenetic trees were generated in MEGA11 (Tamura et al. 2021). Model analysis was used to identify the phylogenetic model with lowest bayesian information criterion.

*GX/P3B*

Datasets GX/P3B and GX/P2V were analyzed using FastQC version 0.11.9 (Andrews, 2014).

We used the following workflow to download and align reads to al mitochondrial genomes on NCBI:
https://github.com/semassey/Scanning-NGS-datasets-for-mitochondrial-and-coronavirus-contaminants/blob/main/complete-mapping-workflow.txt
Bowtie2 v2.4.2 (Langmead and Salzberg, 2012) and samtools v1.14 (Danecek et al. 2021) were used.

Aligned read count histogram was generated by aligning reads to all mitochondrial genomes on NCBI as per above, converting aligned output to fastq format then using textHistogram (https://github.com/ENCODE-DCC/kentUtils) with the following commands:

```
cat aligned.fq | perl -ne '$s=<>;<>;<>;chomp($s);print length($s)."\n";' >
file_out.readslength.txt
textHistogram file_out.readslength.txt -maxBinCount=150
```

Seqkit v2.1.0 (Shen et al. 2016) was used to filter reads to specific minimum lengths.

Dataset GX/P3B was aligned in Galaxy using bowtie2 with the following parameters "--score-min L,0,0" to the human genome (GCF_000001405.39_GRCh38.p13_genomic). Aligned reads were then aligned to the following in sequential order, each time only unaligned reads were kept for the next stage of alignment: ManJav2.0 genomic, ManJav2.0 RNA and ManPten2.0 genomic files. Unaligned reads where then aligned to each of the following genomes in series: Cow, pig, dog using the built-in genome index option for the BowTie2 tool in Galaxy using the "--score-min L,0,0" parameter, again each time only unaligned reads were kept.

The final human only reads were then analyzed using NcbiblastnCommandline local blast using BioPython against the nt database using the following parameters: PER_ID = 80; E_VAL=0.05; MAX_TGT_SEQS=5; MIN_MATCH_LEN=20; MAX_HSPS=1. The nt database was downloaded and built on the 21/1/2021.

All mitochondrial aligned read count and coverage plots were generated using matplotlib v3.3.4 (Hunter 2007).

## Supp Info.

Supplementary information 1, 3, 4 can be accessed at:
doi: 10.5281/zenodo.6807098
Link: Link: https://zenodo.org/record/6807098

Supp Info. 1:
Supp_Info_1_PRJNA641544_DG14_DG18.xlsx

Supp. Info. 2:
A systematic mapping analysis of all mitochondrial genomes present in the NCBI was conducted and the results can be found here:
https://github.com/semassey/Scanning-NGS-datasets-for-mitochondrial-and-coronavirus-contaminants/blob/main/Mito-mappings-DG14-DG18.xlsx

Supp Info. 3:
Supp_Info_3_PRJNA606875_SRR11093270_reads_blast_nt_seq5_hsps1_PCT80_E0.05_hsps.txt

Supp. Info. 4:
Supp_Info_4_PRJNA573298_Analysis.xlsx

# Supp. Figs. and Tables

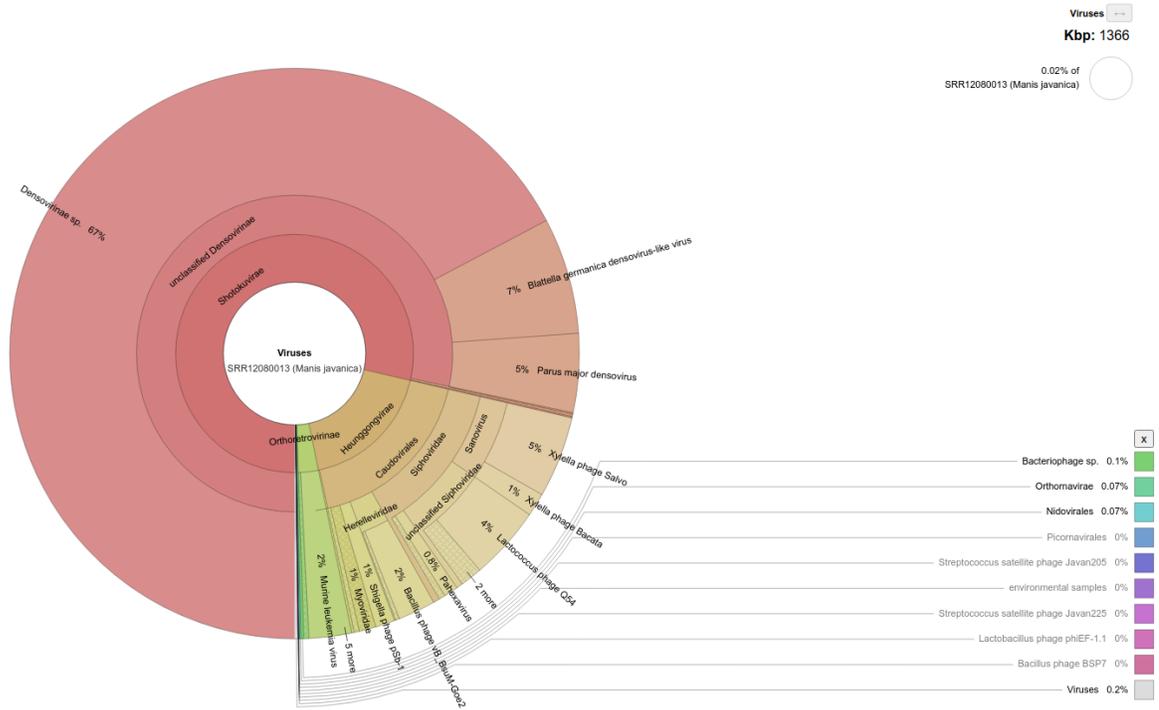

Supp. Fig. 1. STAT Krona virus taxonomy of sample DG14.

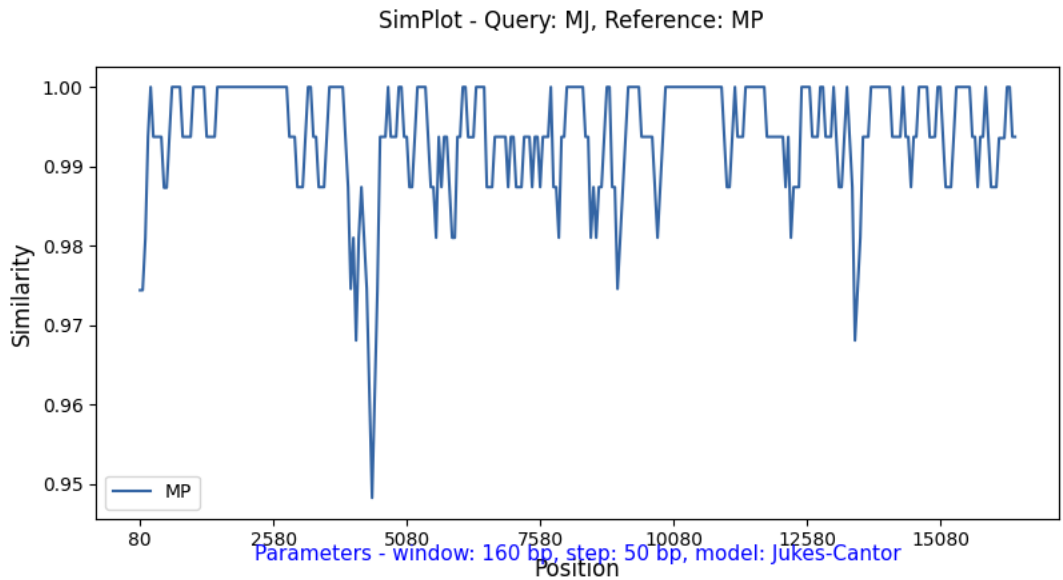

Supp. Fig. 2. SimPlot analysis of *Manis javanica* isolate T298 mitochondrion (MJ) (NC_026781.1) against *Manis pentadactyla* mitochondrion (NC_016008.1).

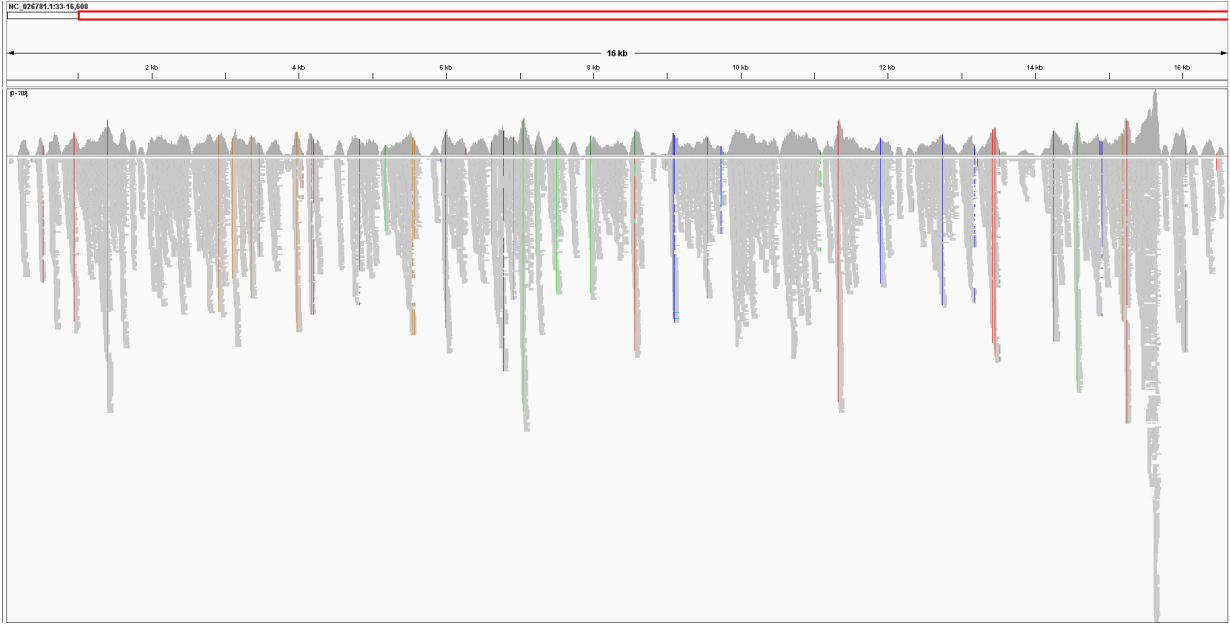

Supp. Fig. 3. Sample DG18 was aligned to the MP mitochondrial genome genome with 100% identity. Aligned reads were then realigned to the MJ mitochondrial genome. 51 SNVs were found (solid vertical lines). All corresponded to SNVs between the MJ and MP mitochondrial genomes.

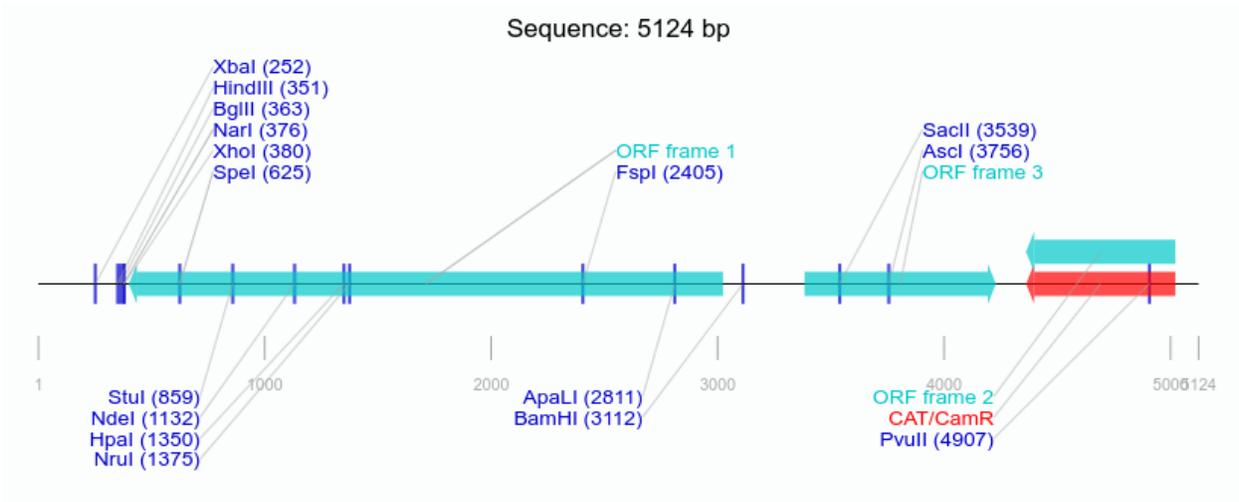

Supp. Fig. 4. Contig k79_60474 after samples DG14 and DG18 were pooled, de novo assembled using MEGAHIT and found to contain a cam repressor (CamR) sequence and multi-cloning site (MCS).

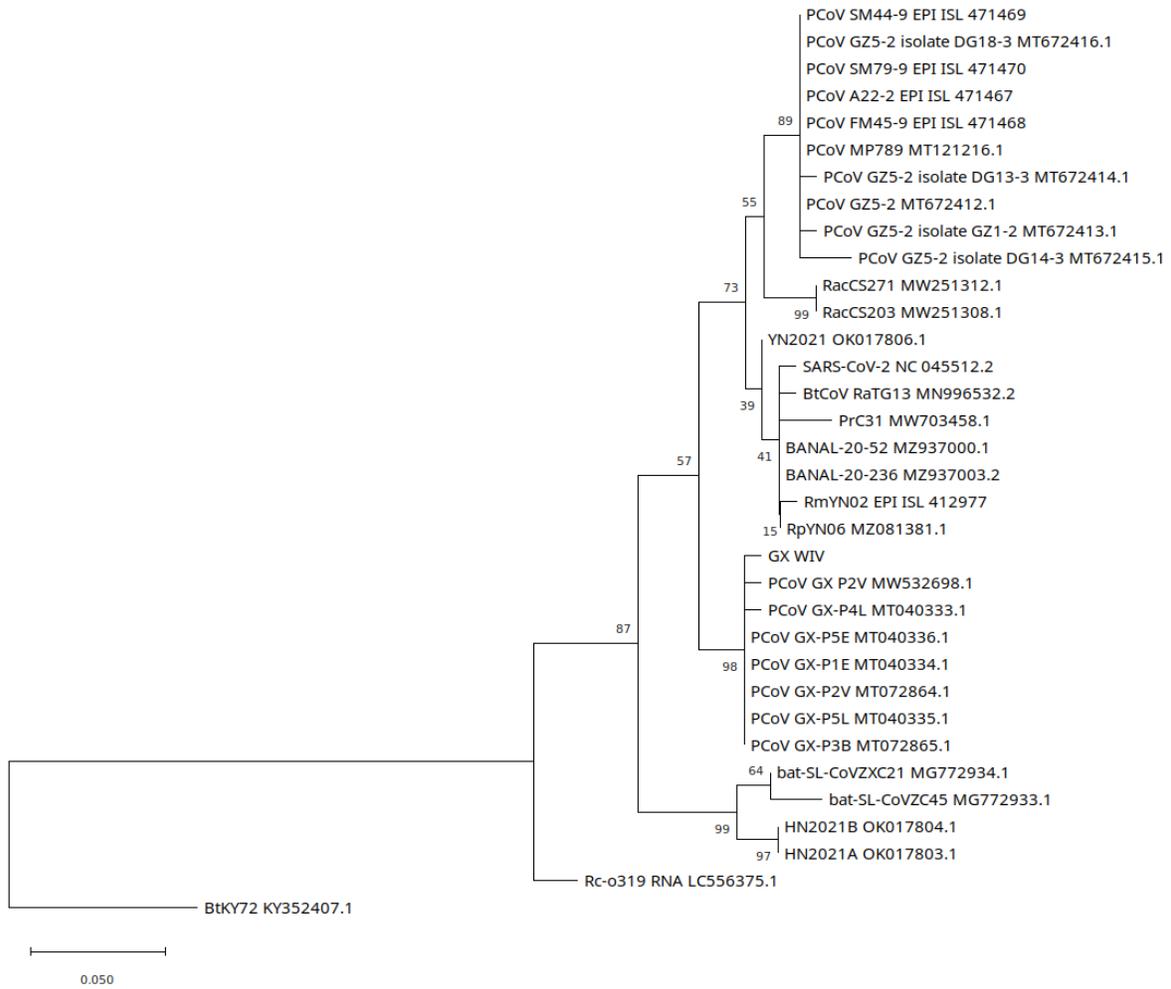

Supp. Fig. 5. Maximum likelihood tree for 160nt section (165nt after multi genome alignment) of N gene for PCoV GZ5-2 isolates GZ1-2, DG13-3, DG14-3, DG18-3 compared with selected SARS2r-CoVs and SRSr-CoVs. Generated using a K2+G model with 1500 bootstrap replicates in MEGA11. Tree rooted on BtKY72.

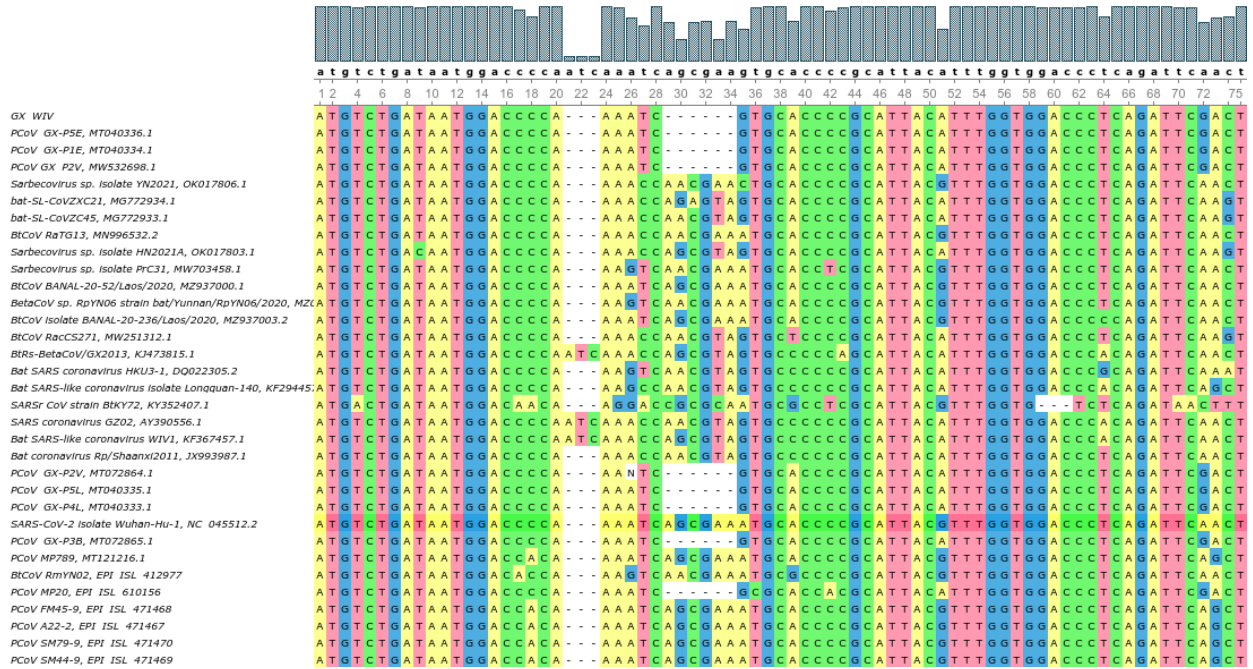

Supp. Fig. 6. Multiple sequence alignment of N gene coding region for SARSr-CoVs. Guanxi PCoVs lack a 6nt sequence near the 5' end of the N gene (26-32nt relative to PCoV MP789, 29-34nt as aligned above).

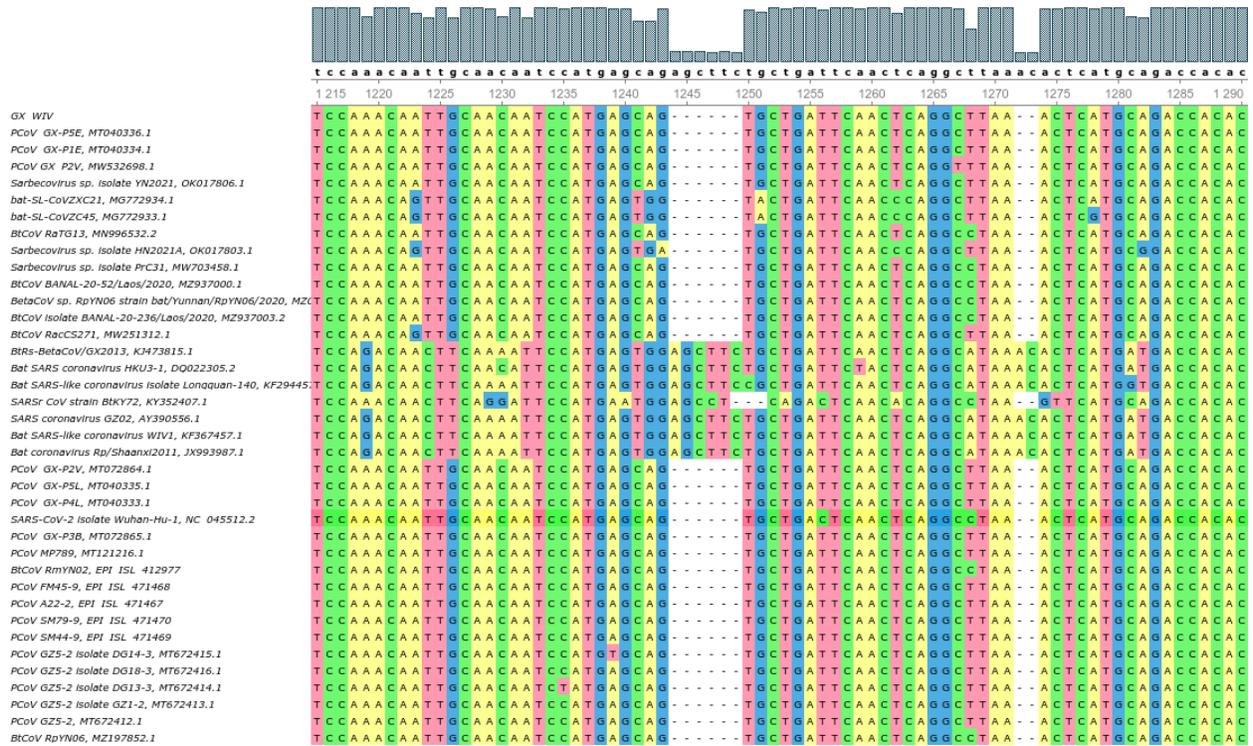

Supp. Fig. 7. Multiple sequence alignment of 3' end of N gene coding region for SARSr-CoVs. Only SARSr-CoVs GX2013, HKU3-1, Longquan-140, GZ02, WIV1 and Shaanxi2011 possess 6nt and 2nt inserts in this region with BtKY72 possessing a single 3nt insert.

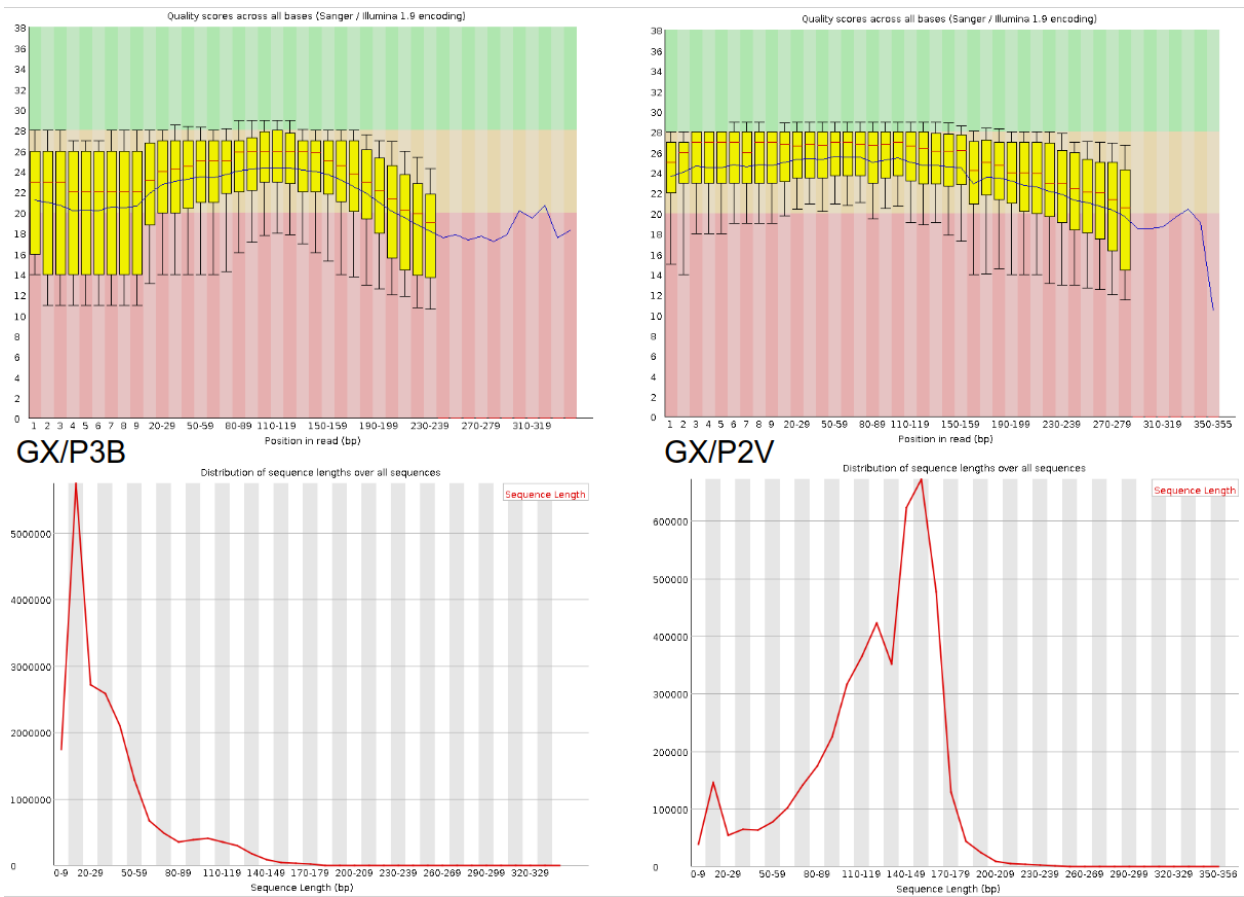

Supp. Fig. 8. FastQC analysis of datasets GX/P3B and GX/P2V showing estimated quality scores and sequence length distributions.

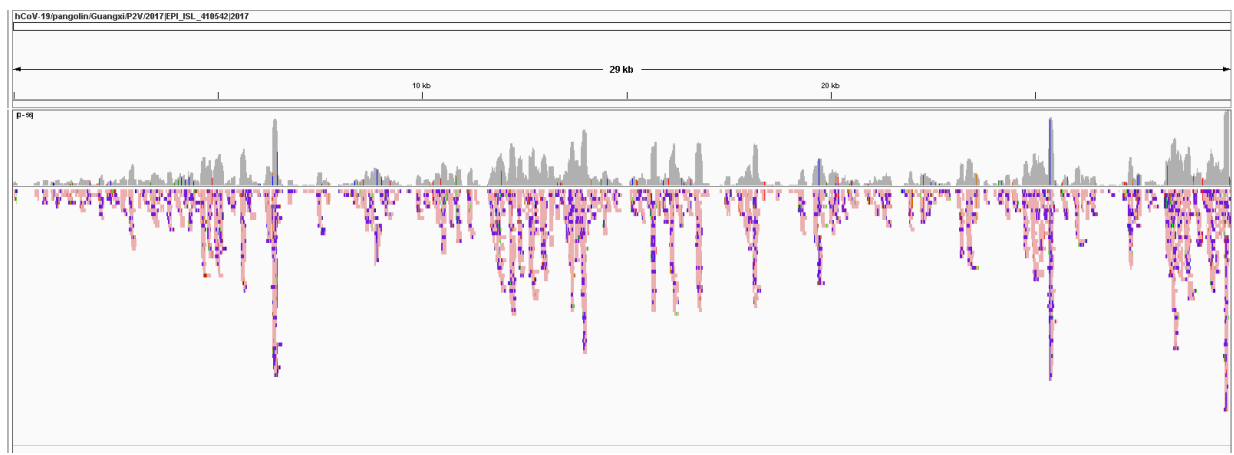

Supp. Fig. 9. GX/P3B (SRR11093270) aligned to PCoV GX-P3B using minimap2 with abundant indels indicated by purple bars.

```
 1  15 ************************************************** 3788
 2  16 ***************************** 2132
 3  17 ******************* 1430
 4  18 ******************** 1446
 5  19 **** 240
 6  20 * 70
 7  21 6
 8  22 2
 9  23 0
10  24 0
11  25 0
12  26 0
13  27 0
14  28 0
15  29 0
16  30 0
17  31 0
18  32 0
```

Supp. Fig. 10. Histogram of read lengths for GX/P3B aligned to the *Picocystis salinarum* mitochondrial genome (NC_042491.1), which exhibited 60% coverage when aligned using the raw GX/P3B dataset. All aligned reads are 22nt length or less indicating spurious alignments.

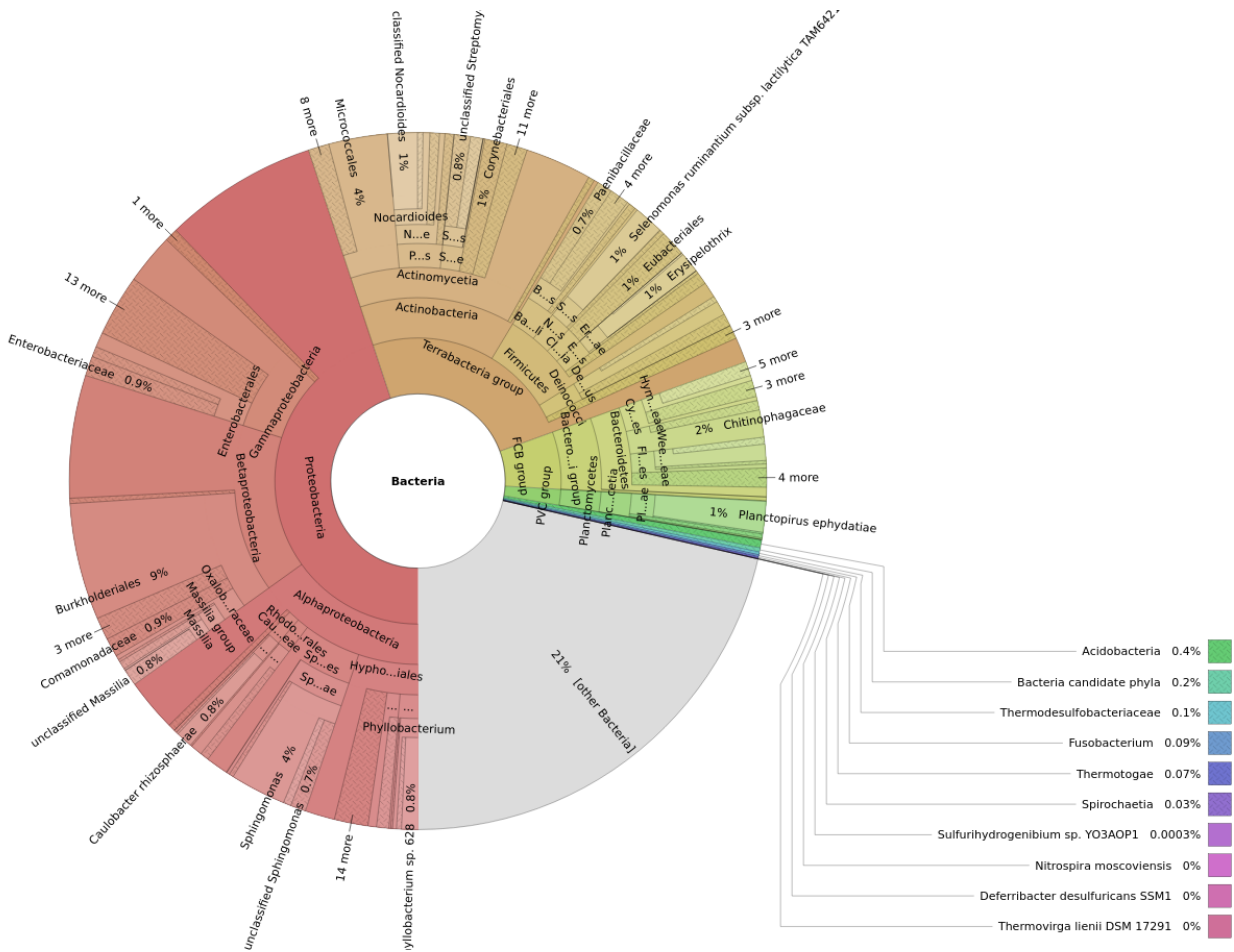

Supp. Fig. 11. Kraken2 analysis of bacterial taxonomy of GX/P3B (SRR11093270), displayed using Krona

Supp. Fig. 12. Kraken2 analysis of Alphaproteobacteria taxonomy of GX/P3B (SRR11093270), displayed using Krona

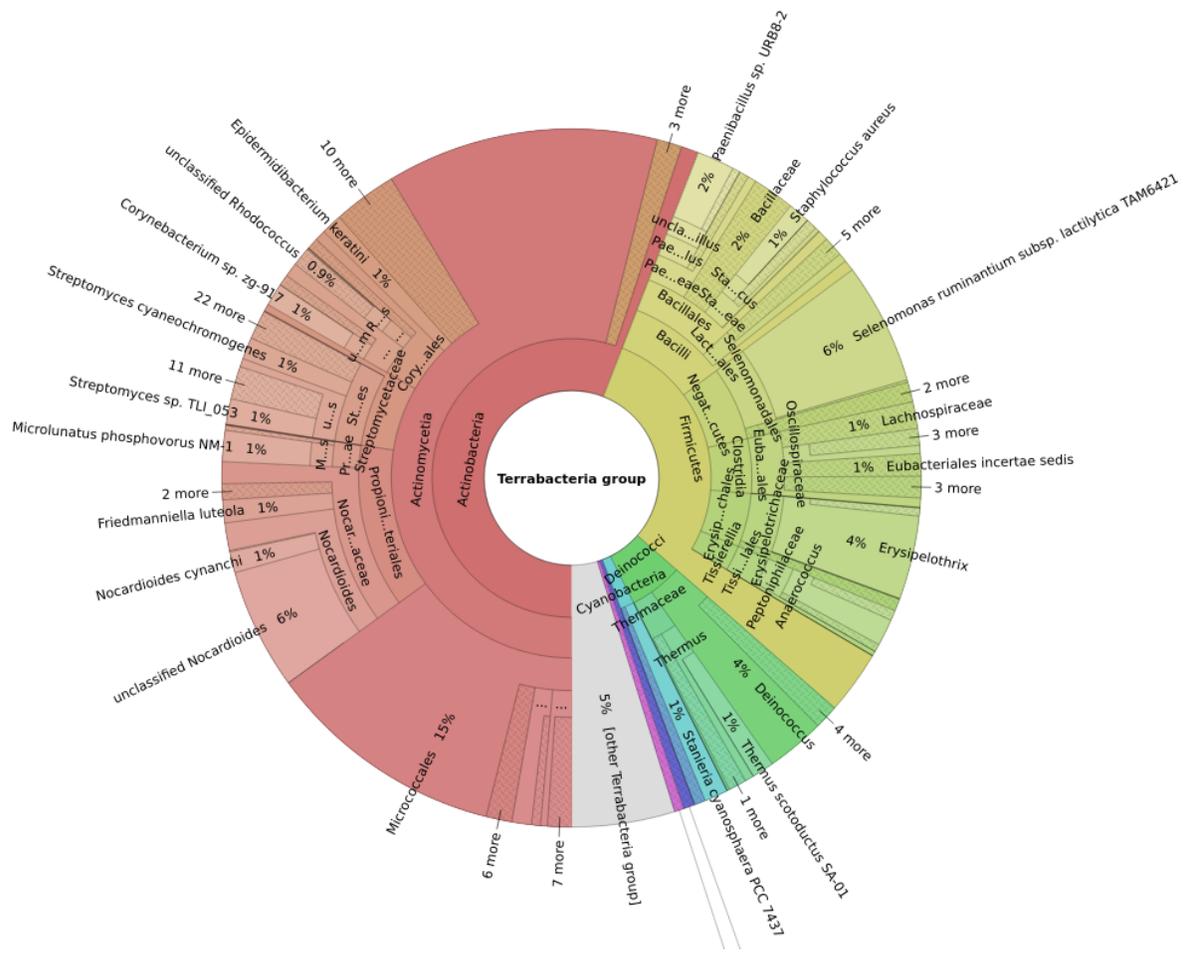

Supp. Fig. 13. Kraken2 analysis of taxonomy of the Terrabacteria group for sample GX/P3B (SRR11093270), displayed using Krona

Supp. Fig. 14. Kraken2 analysis of bacterial taxonomy of GX/P2V (SRR11093271), displayed using Krona

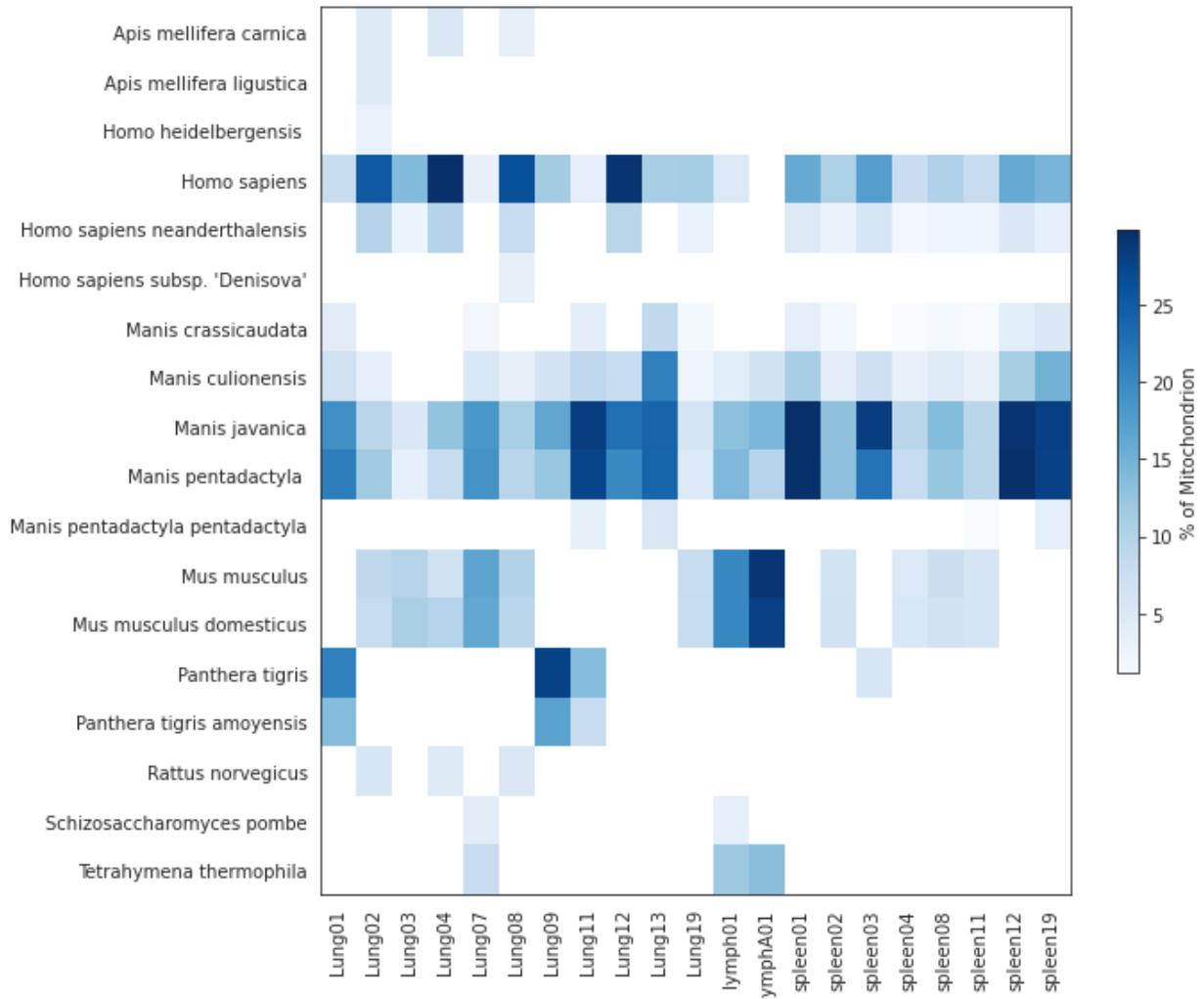

Supp. Fig. 15. Percentage of covered bases for each mitochondrial genome calculated by: summing all covered bases for each mitochondrion with coverage >5% in each dataset then dividing each mitochondrial genome base coverage by the sum.

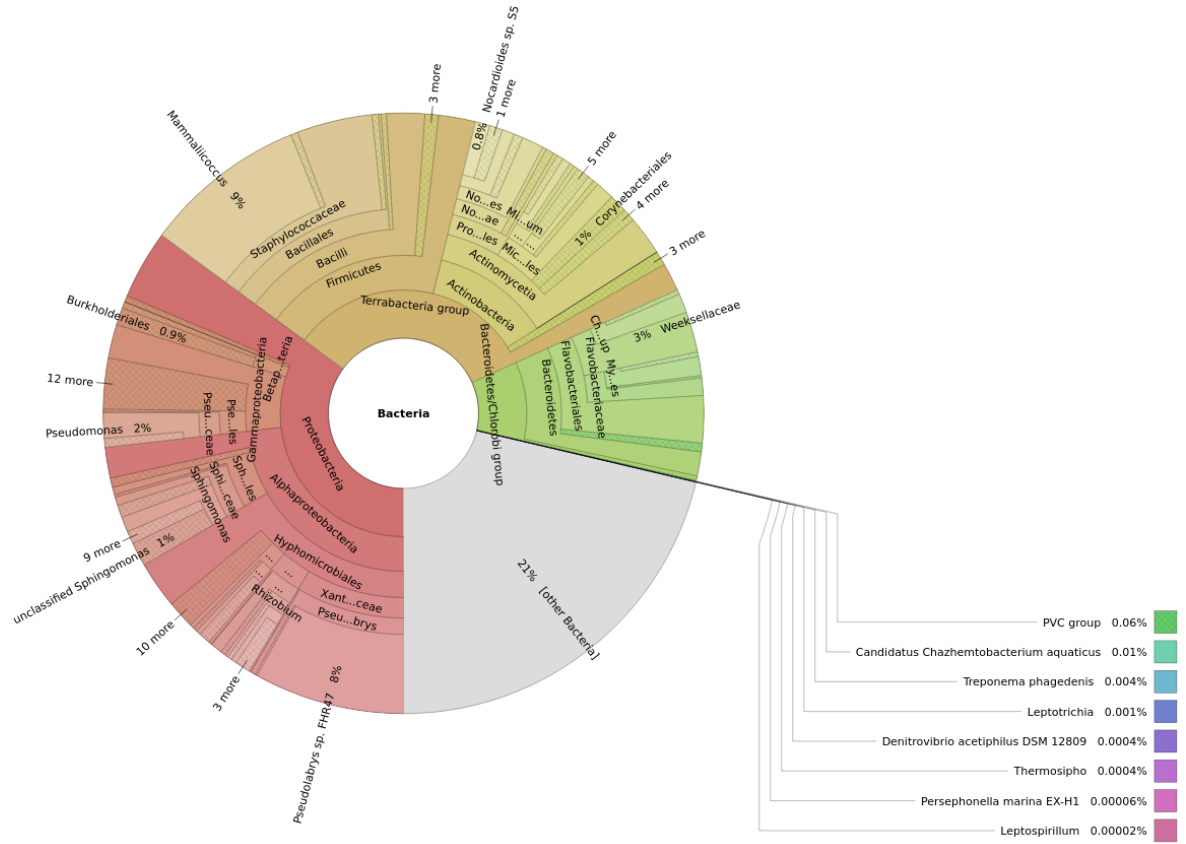

Supp. Fig. 16. Lung09 (SRR10168376) Kraken2 bacterial taxonomic classification using the standard database.

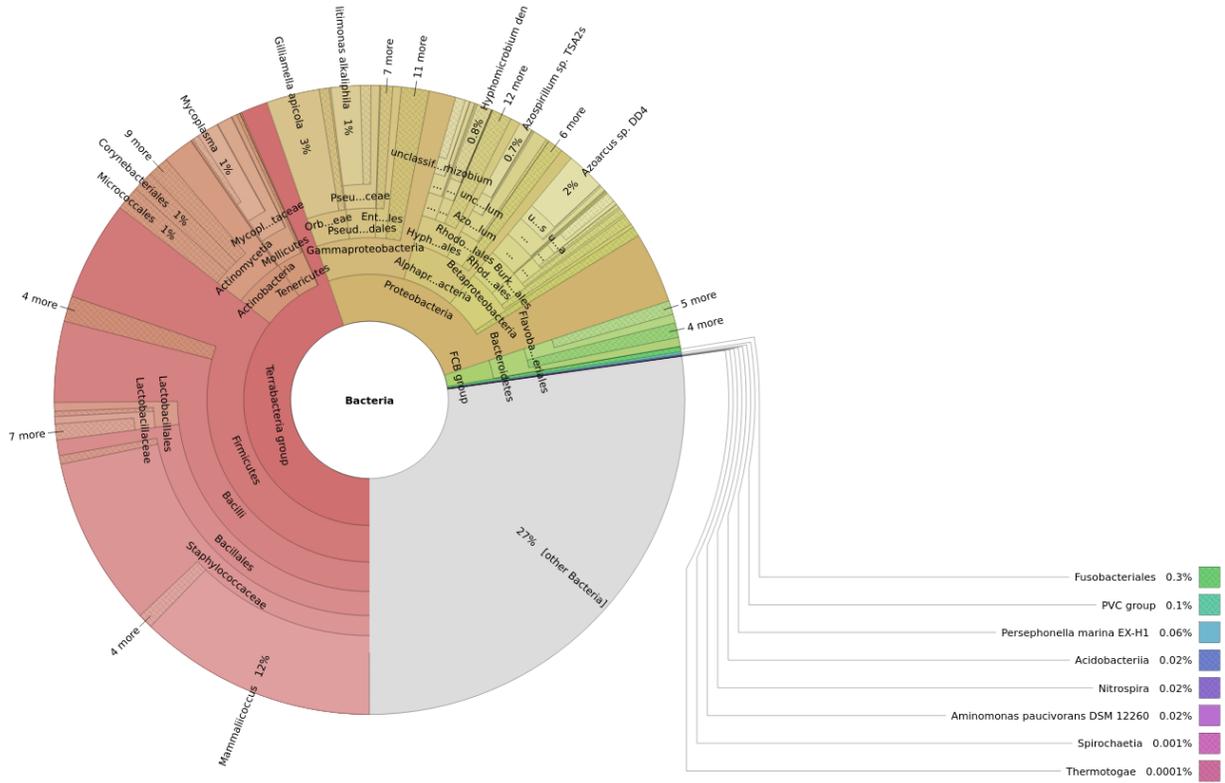

Supp. Fig. 17. Lung08 (SRR10168377) Kraken2 bacterial taxonomic classification using the standard database.

Supp. Fig. 18. Lung07 (SRR10168378) Kraken2 bacterial taxonomic classification using the standard database.

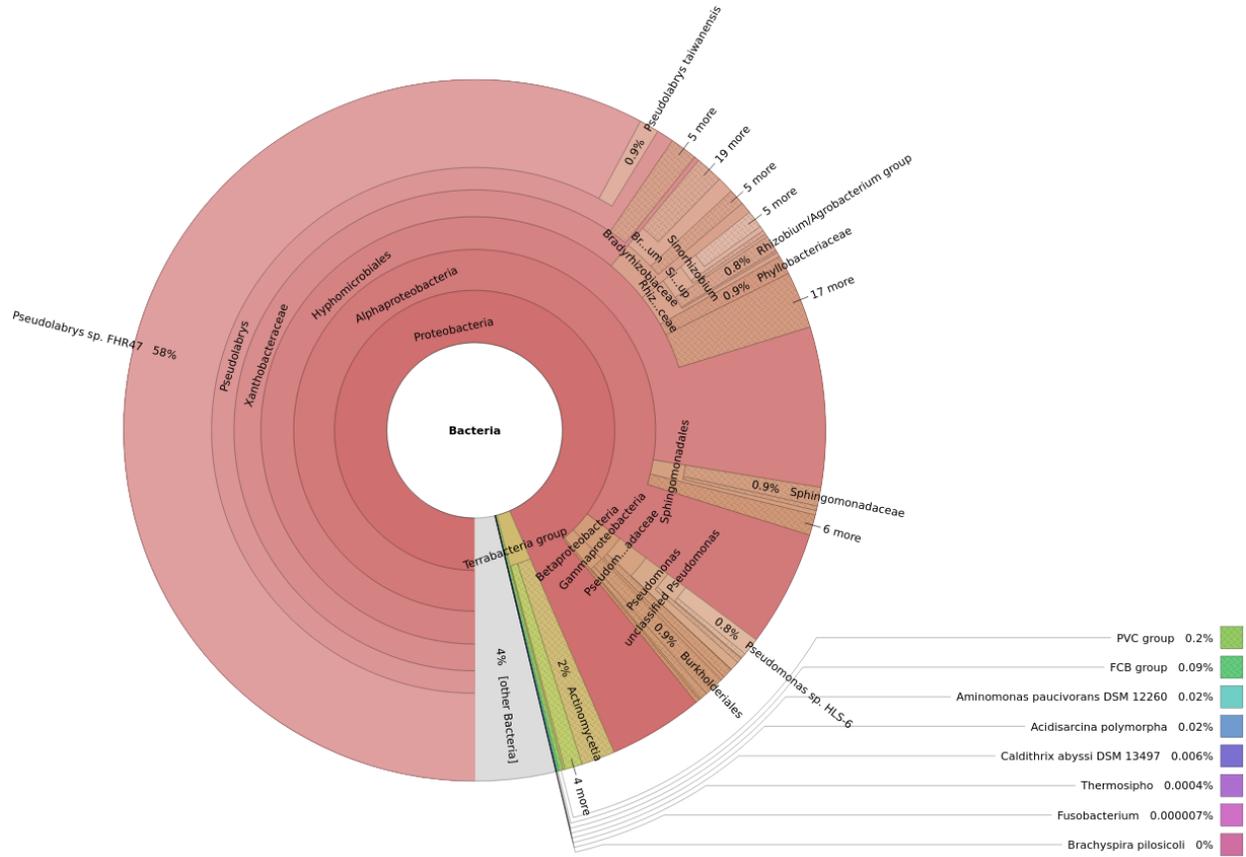

Supp. Fig. 19. Spleen12 (SRR10168382) Kraken2 bacterial taxonomic classification using the standard database.

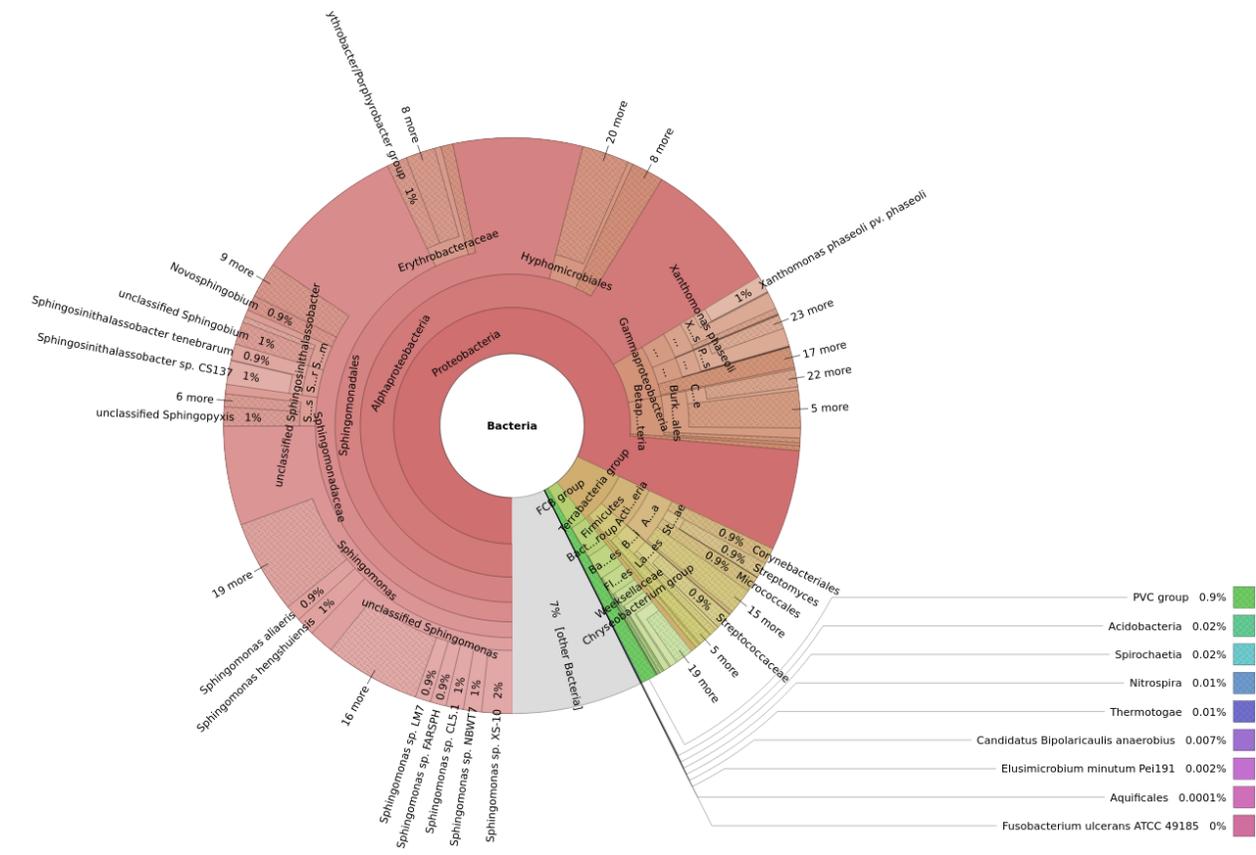

Supp. Fig. 20. Lung13 (SRR10168373) kraken2 bacterial taxonomic classification using the standard database.

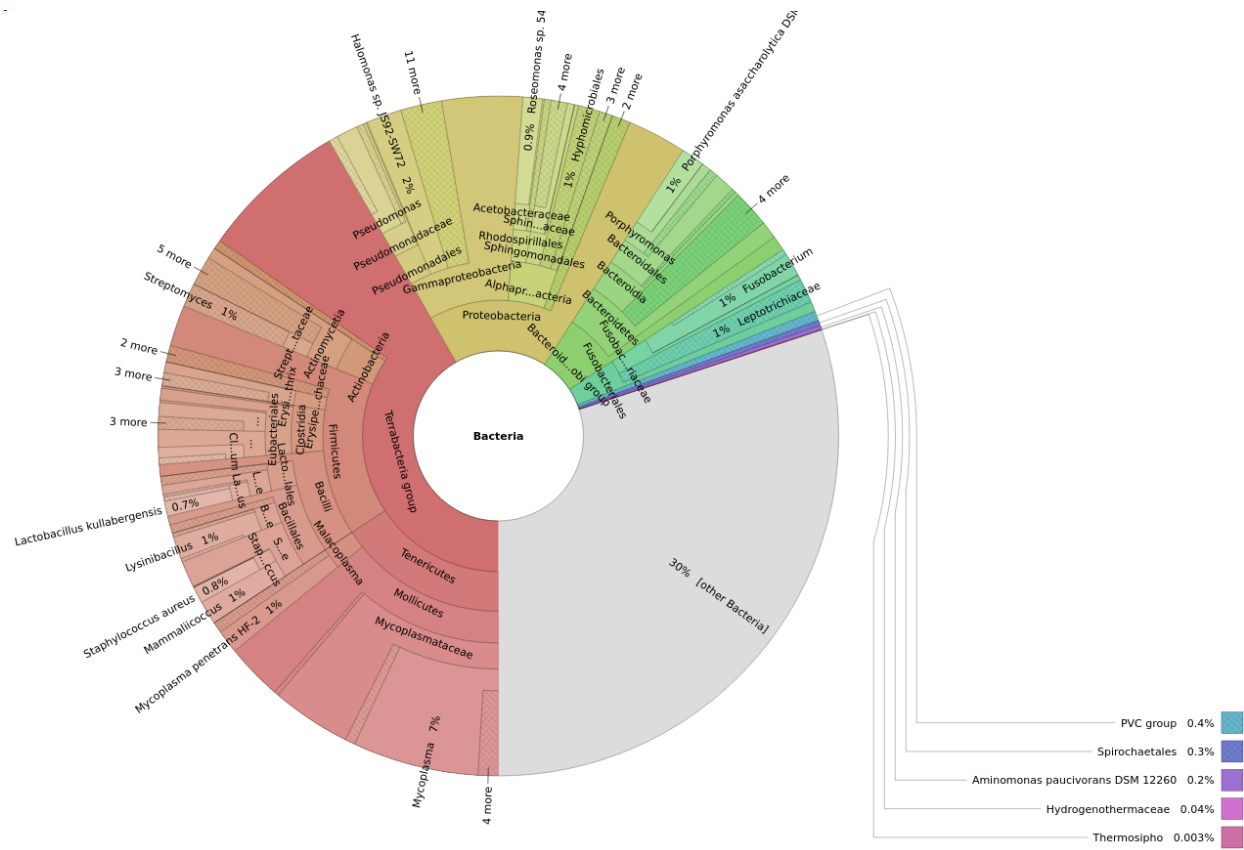

Supp. Fig. 21. Lung12 (SRR10168374) Kraken2 bacterial taxonomic classification using the standard database.

| Mitochondrion | Accession | N | Coverage% | Cov 10x % |
|---|---|---|---|---|
| *Manis pentadactyla* | NC_016008.1 | 15202 | 95.54 | 87.71 |
| *Homo sapiens* | NC_012920.1 | 2714 | 81.32 | 45.87 |
| *Sus scrofa* | MT199606.1 | 17 | 5.37 | 0 |
| *Manis javanica* | NC_026781.1 | 13189 | 95.35 | 85.62 |

Supp. Table 1. SRA dataset DG18 (SRR12080012) alignments to selected mitochondrial genomes with 100% identity

| Mitochondrion | Accession | N | Coverage% | Cov 10x % |
|---|---|---|---|---|
| *Manis pentadactyla* | NC_016008.1 | 17079 | 93.35 | 82.43 |
| *Homo sapiens* | NC_012920.1 | 485 | 66.92 | 3.52 |
| *Sus scrofa* | MT199606.1 | 100 | 13.32 | 0.45 |
| *Manis javanica* | NC_026781.1 | 24552 | 99.01 | 96.21 |

Supp. Table 2. SRA dataset DG14 (SRR12080013) alignments to selected mitochondrial genomes with 100% identity

| id | title | accession | length | query_length | score | expect value |
|---|---|---|---|---|---|---|
| k141_6348_335 | Cloning vector pMSCV-syn-Gephyrin.FingR-GFP | MT612434 | 7231 | 335 | 332 | 1.37E-171 |
| k141_283_106 | Cloning vector pXJ070 | MT084772 | 7170 | 106 | 106 | 1.43E-46 |
| k141_4825_662 | Cloning vector pBAD24-rnc-sfGFP | KF020495 | 5932 | 662 | 560 | 0 |
| k141_3863_644 | PCoV_GX-P2V | MT072864 | 29795 | 644 | 508 | 0 |
| k141_6635_619 | PCoV_GX-P2V | MT072864 | 29795 | 619 | 561 | 0 |
| k141_266_342 | PCoV_GX-P2V | MT072864 | 29795 | 342 | 342 | 3.87E-177 |
| k141_4201_443 | PCoV_GX-P3B | MT072865 | 29801 | 443 | 440 | 0 |
| k141_5354_804 | PCoV_GX-P2V | MT072864 | 29795 | 804 | 801 | 0 |
| k141_870_480 | PCoV_GX-P2V | MT072864 | 29795 | 480 | 447 | 0 |
| k141_4452_416 | PCoV_GX-P3B | MT072865 | 29801 | 416 | 416 | 0 |
| k141_3326_597 | PCoV_GX-P5E | MT040336 | 29802 | 597 | 591 | 0 |
| k141_4198_402 | PCoV_GX-P3B | MT072865 | 29801 | 402 | 398 | 0 |
| k141_5497_469 | PCoV_GX-P3B | MT072865 | 29801 | 469 | 460 | 0 |
| k141_1561_657 | PCoV_GX-P3B | MT072865 | 29801 | 657 | 654 | 0 |
| k141_6593_471 | PCoV_GX-P3B | MT072865 | 29801 | 471 | 468 | 0 |
| k141_2142_952 | PCoV_GX-P3B | MT072865 | 29801 | 952 | 919 | 0 |
| k141_4148_702 | PCoV_GX-P3B | MT072865 | 29801 | 702 | 674 | 0 |
| k141_2776_316 | PCoV_GX-P3B | MT072865 | 29801 | 316 | 295 | 4.75E-151 |
| k141_1675_401 | PCoV_GX-P3B | MT072865 | 29801 | 401 | 396 | 0 |
| k141_283_384 | Cloning vector pColE-AT | MN623119 | 7664 | 384 | 137 | 3.99E-63 |

Supp. Table 3. Dataset GX/P3B (SRR11093270) was *de novo* assembled using MEGAHIT, and contigs were aligned to a combined NCBI univec & viral database using minimap2, then aligned contigs were analyzed using blast against a local copy of the nt database.